# Mie resonance engineering in meta-shell supraparticles for nanoscale nonlinear optics

Joong Hwan Bahng[1, ‡], Saman Jahani[1,‡], Douglas Montjoy[2], Timothy Yao[1], Nicholas Kotov[2], Alireza Marandi[1,*]

[1] Department of Electrical Engineering, California Institute of Technology

[2] Department of Chemical Engineering, University of Michigan, Ann Arbor

* email: marandi@caltech.edu

**Abstract**

Supraparticles are coordinated assemblies of discrete nanoscale building blocks into complex and hierarchical colloidal superstructures. Holistic optical responses in such assemblies are not observed in an individual building block or in their bulk counterparts. Furthermore, subwavelength dimensions of the unit building blocks enable engraving optical metamaterials within the supraparticle, which thus far has been beyond the current pool of colloidal engineering. This can lead to effective optical features in a colloidal platform with unprecedented ability to tune the electromagnetic responses of these particles. Here, we introduce and demonstrate the nanophotonics of meta-shell supraparticle (MSP), an all dielectric colloidal superstructure having an optical nonlinear metamaterial shell conformed onto a spherical core. We show that the metamaterial shell facilitates engineering the Mie resonances in the MSP that enable significant enhancement of the second harmonic generation (SHG). We show several orders of magnitude enhancement of second-harmonic generation in an MSP compared to its building blocks. Furthermore, we show an absolute conversion efficiency as high as $10^{-7}$ far from the damage threshold, setting a new benchmark for SHG with low-index colloids. The MSP provides pragmatic solutions for instantaneous wavelength conversions with colloidal platforms that are suitable for chemical and biological applications. Their engineerability and scalability promise a fertile ground for nonlinear nanophotonics in the colloidal platforms with structural and material diversity.

**Graphical Table of Contents Entry**

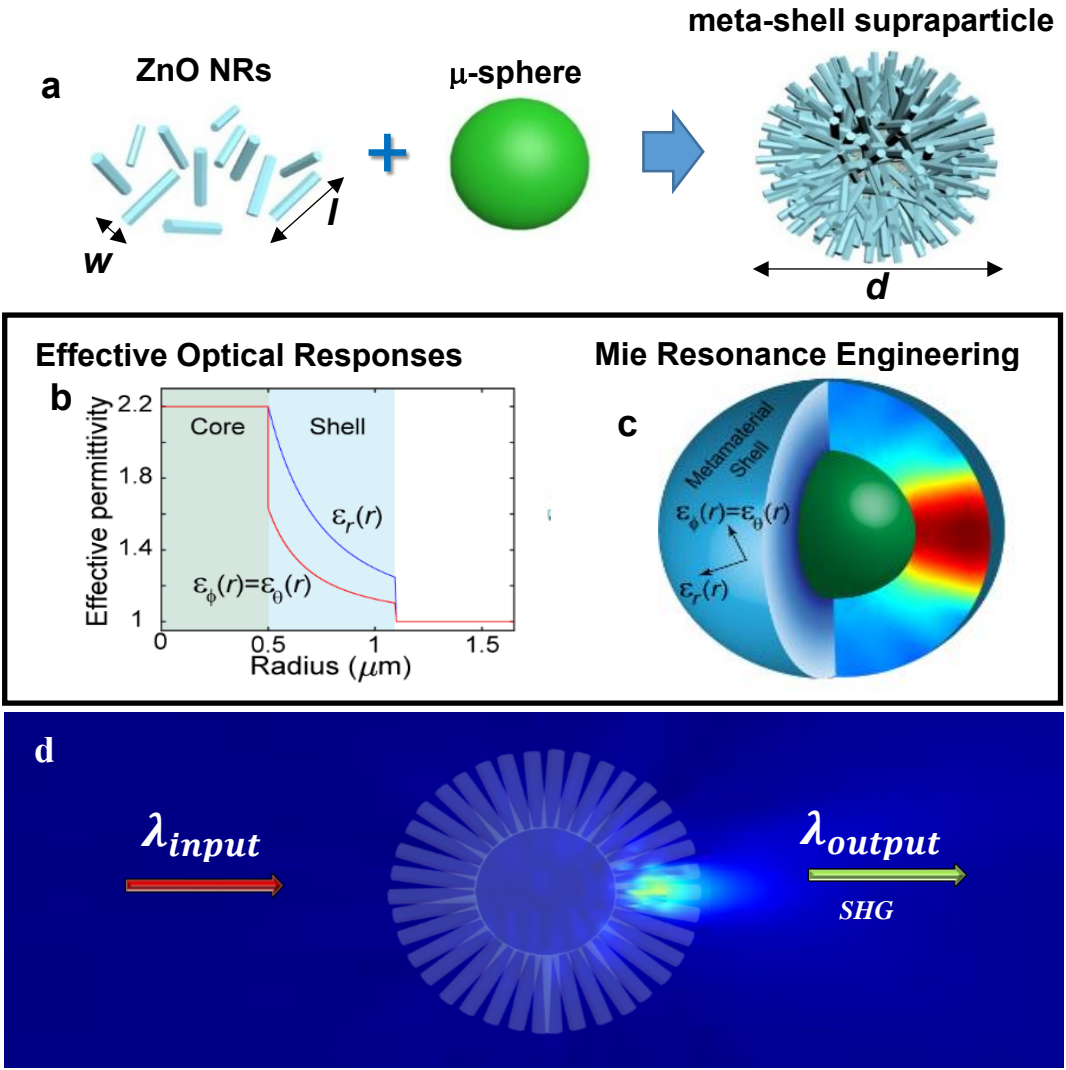

**Introduction**

Colloidal nanoparticles (CNPs) and their assemblies exhibit a cohort of electromagnetic (EM) resonances and couplings, facilitating them as essential optical actuating platforms in bio/-chemical sensing, imaging and photo-catalytic reactions[1–6]. Their processing versatility enable easy integration into photonic and solid-state devices[7,8] as modular light sources. When assembling the colloidal nanoparticles into colloidal superstructures, holistic optical responses are observed that are not found in an individual building block or in their bulk counterparts[9]. Notwithstanding, CNPs that enable instantaneous and efficient nonlinear harmonic conversion of electromagnetic (EM) stimuli could provide solutions to the remaining important challenges in their photo-activation and -actuation. For example, wavelength requirements to trigger a photo-stimulation could bring limitations due to insufficient light penetration into the reaction media and biological tissues[10,11]. Delivering optical energy through the ambient media within its specific transparency window that could trigger the target photo-responses via instantaneous wavelength conversions could immensely broaden the scope of the operation while enhancing its procedural compatibilities and efficiencies[12–15].

Recently, several advanced strategies have been employed to enhance optical wavelength conversions in nanostructures. A large subset of such strategies are focused on high index materials that are architected via top-down nanofabrication[16–23]. Synthetic challenges exist in realizing dexterous colloidal platforms with high index materials and their bottom-up fabrication. As such, communities have explored low index nano-colloids for optical nonlinear conversions[24–28]. However, limitations in their structural diversities and complexities constrains their engineerability for nonlinear nanophotonics.

Here, we demonstrate a pragmatic approach to enhance optical nonlinear conversions with low index $\chi^{(2)}$ CNPs by assembling them into a colloidal superstructure in the form of a synthetic metamaterial shell conformed onto a dielectric core sphere, which can be denoted as the meta-shell supraparticles (MSP). In comparison to individual $\chi^{(2)}$ CNPs, their assembly into an MSP increases the density of Mie resonances and their combinations for $\chi^{(2)}$ nonlinear optical interactions. The optical response of the metamaterial shell in these particles can be utilized for engineering the collective behavior of these Mie resonances, allowing for a significant enhancement in the 2nd order nonlinear optical wavelength conversion.

**Results and Discussions**

The MSP is realized through one-pot chemical synthesis (**SI 1.1**). We have previously demonstrated hierarchical colloidal superstructures with remarkable assembly quality and yield, using multiple materials selections with various constitutive properties, shapes, and dimensions [29–31]. Such a synthetic versatility enabled us to engineer structures via simple procedural adjustments. The MSP can have varieties of core spheres with tailored meta-shell nano-topography comprised of vertically oriented ZnO nano rods (NRs) as the non-centrosymmetric $\chi^{(2)}$ material[32] (a few examples in **Figure 1d-e** (**SI 1.1**)).

First, we carried out linear light scattering measurements by the MSPs in aqueous dispersion and realized that their extinction lineshapes do not correlate with either the ZnO NRs or the core μ-spheres, **Figure 1f**. The spectra feature a broadband peak ($P_1$) in the visible wavelengths and a narrower peak ($P_2$) in the ultraviolet range. When $P_2$ is approximately at the second harmonic of the $P_1$, the overlap of the EM modes at both the fundamental and the second-harmonic wavelengths suggests opportunities for wavelength conversion. This emergent spectral behavior arising from multitude of Mie resonances can be fine-tuned by adjusting the meta-shell corrugation geometry or its core dimensions, **Figure 1g – i**. These experimental findings on the linear response of the MSPs are verified with finite difference time domain (FDTD) full-wave simulations. We created a model MSP (SI 2.2) that mirrors the experimental construct based on a silica core sphere with an overall diameter of $d = 2.2$ μm (**Figure 1d**), i.e. MSP2.2. The model replicates the imperfect orthogonal orientation of the ZnO NRs. Numerical results of extinction cross-section $\sigma_{ext}$ of MSP2.2 are in agreement with the experimental measurements, as depicted in **Figure 1j, k** (**SI 2.2 – 2.4**).

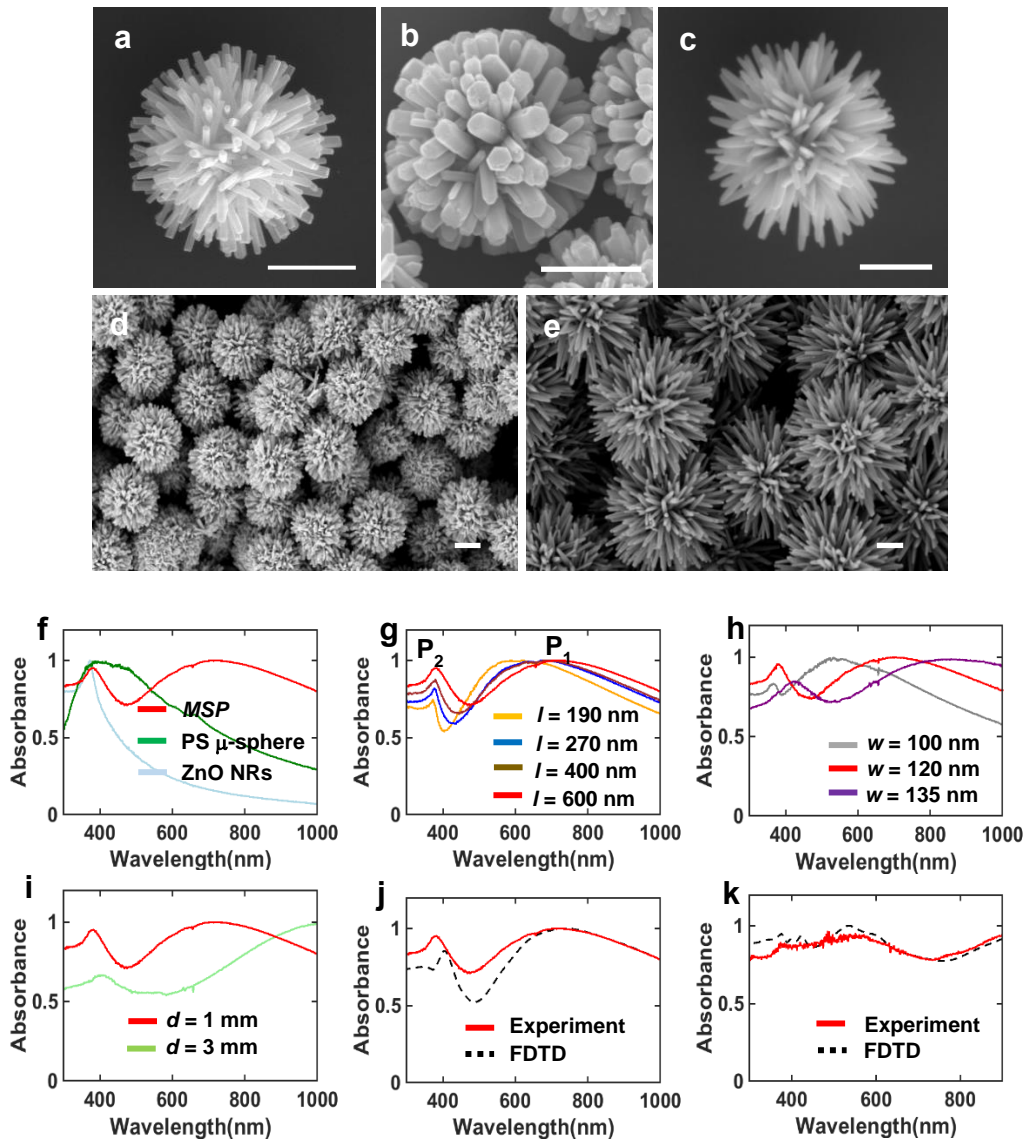

**Figure 1. Structural engineering of MSP. a-c,** scanning electron microscopy (SEM) images of the MSP synthesized with a polystyrene (PS) μ-sphere core (diameter, $d$ = 1 μm) having diverse geometries and dimensions of its unit building block ZnO NRs that constitute the meta-shell; **d, e,** MSP synthesized with a SiO$_2$ μ-sphere template ($d$ = 1 μm) and having ZnO NR (**d**) length $l \approx$ 600 nm, designated MSP2.2 and (**e**) $l \approx$ 1290 nm, designated MSP3.58; **f-i,** spectral tuning is achieved by adjusting the corrugation geometry in the meta-shell, (**f**) peak normalized (P.N.) extinction spectra (A.U.) , in aqueous dispersion, of ZnO NRs (—), $d$ = 1 μm PS μ-sphere (—) and MSP (—); P.N. extinction spectra of MSP with (**g**) varying spike lengths, $l \approx$ 190 nm (—), $l \approx$ 270 nm (—), $l \approx$ 400 nm (—), $l \approx$ 600 nm (—) and (**h**) varying spike widths, $w \approx$ 100 nm (—), $w \approx$ 120 nm (—), $w \approx$ 135 nm (—), and (**i**) varying core diameters, $d$ = 1 μm (—), $d$ = 3 μm (—); **j-k,** Overlap in the spectral lineshape between the extinction cross-section ($\sigma_{ext}$) of a model MSP from the FDTD full wave simulation (•••), and the extinction spectra from the experimental measurement (—), for both suspended in (**j**) water and in (**k**) air. All scale bar: 1 μm

The linear response of the meta-shell supraparticle, which consists of a metamaterial shell on a core, is enriched by the density of its Mie resonances. While ED, EO and MD modes are supported in the core-only silica sphere ($d_{core}$ = 1 μm) at λ = 1550 nm, our theoretical analysis suggests that the meta-shell particle accommodates electric -dipole (ED), -quadrupole (EQ), -octupole (EO), -hexadecapole (EH) and magnetic -dipole (MD), -quadrupole (MQ), -octupole (MO) modes (**SI 4.2**), as shown in **Figure 2**. These Mie resonances and their interferences can be tuned in the MSP to lead to a hot spot with strong electric field strength |E|. Such a behavior is similar to formation of a photonic nanojet in simple spherical geometries[33,34], however, a significant portion of this hot spot forms in the shell and its |E| strength can be much larger than that of the core alone (see an example with two-fold |E| enhancement, **SI 3.2**).

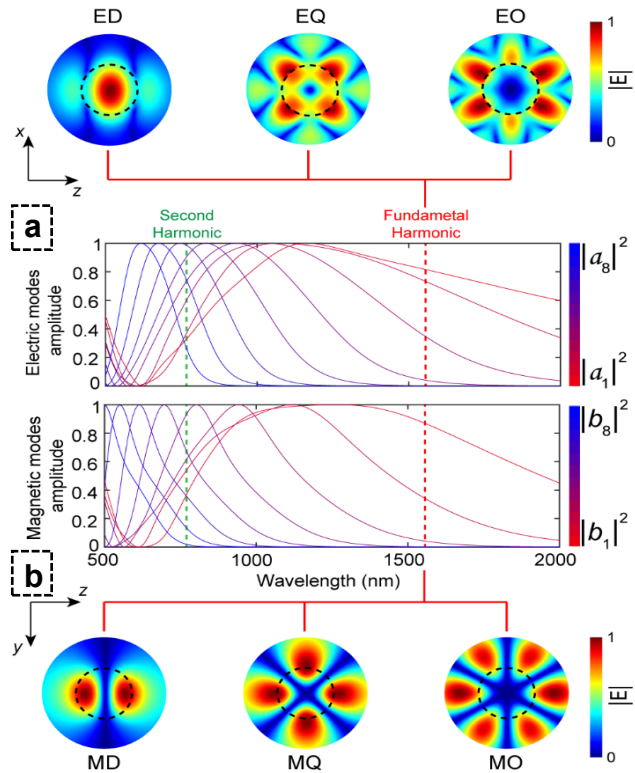

**Figure 2. Mie resonance enrichment with *meta-shell*.** Scattering coefficients of (**a**) electric and (**b**) magnetic multipoles in the MSP calculated from analytical modeling and Mie theory; compared to a core μ-sphere, the density of Mie resonances is enriched in the presence of a circumambient meta-shell. See **SI 4.2** for scattering coefficients corresponding core sphere. Here, $a_n$ and $b_n$ are the scattering coefficients of electric and magnetic dipole in the $n^{th}$ order, respectively.

Engineering the Mie resonances of the MSP and their interferences can be achieved through tuning the radial graded-index profile as well as the angular anisotropy of the metamaterial shell (**Figure 3**). The key distinction of the metamaterial shell compared to the conventional metamaterials is its spherical construct of high aspect-ratio unit building blocks. Such arrangement leads to a radial graded-index profile. Our computational analysis (**SI 4.1**) shows that changing the graded-index profile enables us to change the scattering behaviors of the particle at both the fundamental and the second harmonic wavelengths, **Figure 3a-c**. Moreover, the effective index of the shell near the periphery of the particle (due to lower ZnO NR density) generates an excellent impedance-matched interface which can help to reduce the backscattering of the incident light at the interface, leading to improvement in the light delivery to the shadow side of the supraparticle. As the graded-index tapers to free space, maximal spectral overlap in the ED and the MD mode is achieved across a broad spectrum, **Figure 3c (SI 4.3)**. This can facilitate spatial overlap of the photonic nanojet at both the fundamental and the second-harmonic wavelengths, **Figure 3d**. Improvement in the spatial and spectral overlaps can lead to enhancement of the second-harmonic generation in the photonic nanojet, and its efficient forward propagation, as shown in **Figure 3e, f**. Hence, the radial graded index profile in the MSP can be instrumental in enhancing the SHG and its directionality that is suitable for applications in chemical and biological settings.

The spherical arrangement of orthogonal ZnO NR array in the metamaterial shell also leads to angular anisotropy in the spherical coordinate, which does not appear in natural dielectrics. Such an anisotropy is expected to provide additional degrees of freedom in the overall optical responses [35–37]. In the metamaterial shell, the spherical anisotropy enables fine-tuning the spatial location of the photonic nanojet hotspot. This capability can be utilized for enhancing the nonlinear conversion efficiency by overlapping the hot spot with the highest densities of $\chi^{(2)}$ nanostructures. Increasing the anisotropy (decreasing the $\varepsilon_\theta$, while keeping $\varepsilon_r$ constant) can move the hotspot towards the core interface, **Figure 3g-i**, hence contributing to the enhancement of the nonlinear conversion process.

Our analysis based on the Mie theory considering the graded index profile and the spherical anisotropy of meta-shell (**SI 4.1, 4.2**) matches the full-wave FDTD simulations and explains the superior nonlinear performance of the meta-shell supraparticles (**SI 3.2**). Without the meta-shell, the $SiO_2$ core sphere generates a photonic nanojet whose region of peak $|E|$ is confined within the core sphere, **Figure 3j**. In the presence of a meta-shell featuring a graded index profile only, the peak $|E|$ at the hotspot region is enhanced by an approximate 2-fold, but localized at the outer peripheries, **Figure 3k**. While the peak $|E|$ is higher without the spherical anisotropy in the meta-shell, ascribing spherical anisotropy to the meta-shell fine-tunes the spatial location of the hotspot volume towards, but exterior to, the core interface, at which the density of the $\chi^{(2)}$ nanostructures are at its highest, **Figure 3l**. Overall, the simulations show an approximate 4-fold increase in the peak $|E|$ at the hotspot in the MSP compared to that of the incident light $|E|$ (**SI 3.3, Figure S8**).

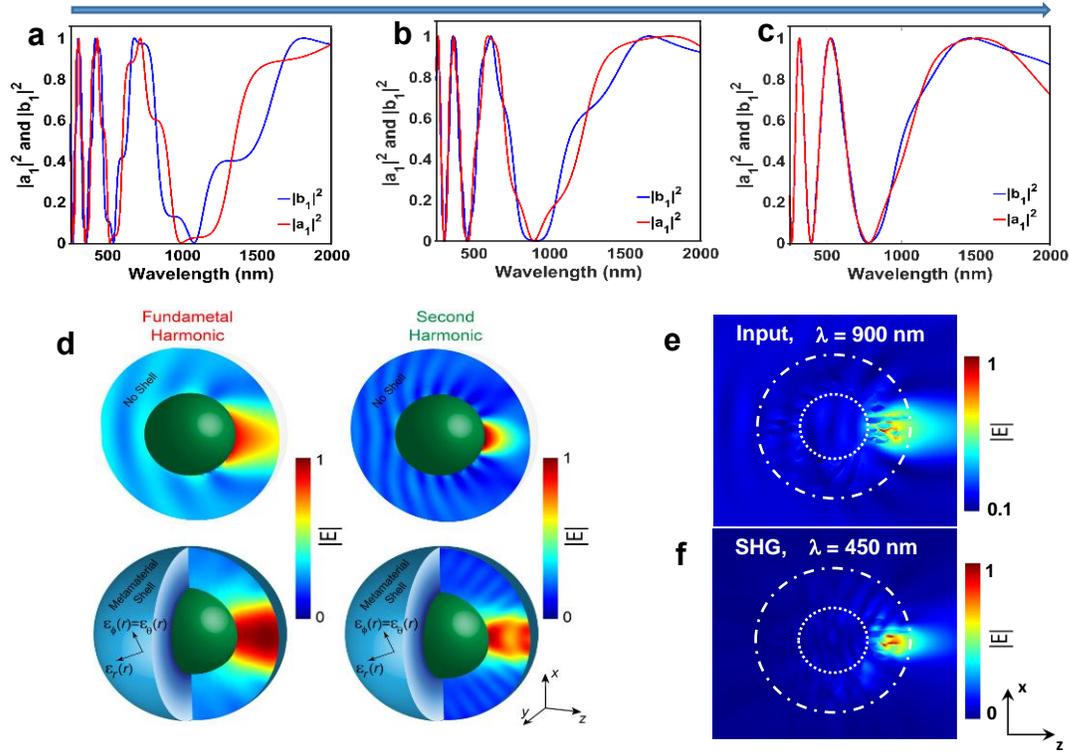

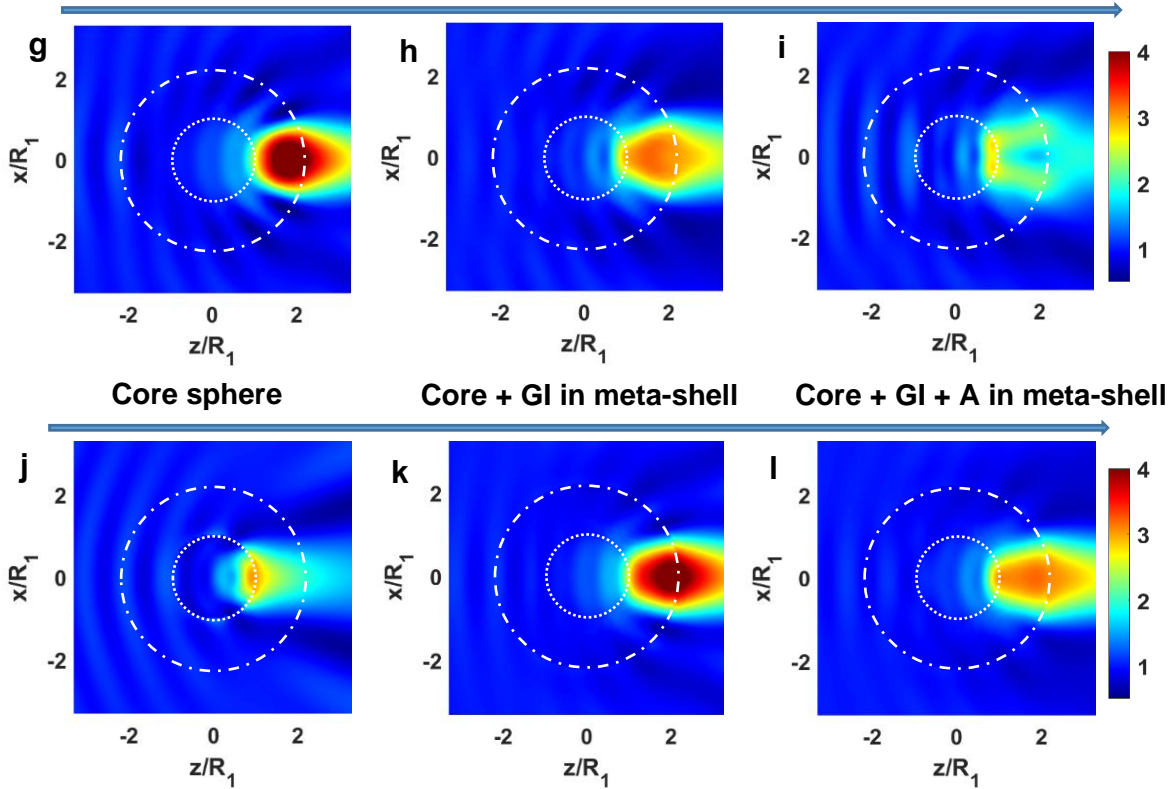

**Figure 3. Mie resonance engineering with meta-shell.** The diameter of the core sphere is $d_{core}$ = 1 μm, thickness of the meta-shell is $t_{shell}$ = 600 nm. The permittivity of the sphere is $\varepsilon_{core}$ = 2.2; **a-c,** increase in the index gradient increases the broadband spectral overlap between the electric dipole (ED) and the magnetic dipole (MD) mode that spans several higher order harmonics, (**a**) $\varepsilon_{r,\,shell}$ : 2.2, (**b**) $\varepsilon_{r,\,shell}$ : 2.2 → 1.5, (**c**) $\varepsilon_{r,\,shell}$ : 2.2 → 1; $a_1$ and $b_1$ are the scattering coefficients of electric and magnetic dipole in the first order, respectively. (**d**) Multi-mode interferences leading to the formation of photonic nanojet hotspot and their spatial overlap at both fundamental and the second harmonic wavelengths; **e-f**, FDTD full wave simulation, at λ = 900 nm, showing (**e**) photonic nanojet formed by the model MSP2.2 upon light incidence and (**f**) enhanced forward scattering in the SHG radiation pattern by the model MSP, alike to a photonic nanojet. The details of the nonlinear optics simulations with the FDTD can be found in the **SI 5 - 7**; **g-i**, Increase in the angular anisotropy shifts the hotspot towards the core sphere interface, (**g**) $\varepsilon_{\theta,\,shell}$ : 2.2, (**h**) $\varepsilon_{\theta,\,shell}$ : 1.5, (**i**) $\varepsilon_{\theta,\,shell}$ : 1; Here, the radial anisotropy is kept constant, $\varepsilon_{r,\,shell}$ = 2.2; **j-l**, photonic nanojet features, calculated from the analytical modeling, for (**j**) the core sphere, (**k**) core sphere with meta-shell having gradient index (GI) feature and (**l**) core sphere with meta-shell having both the gradient index and anisotropy (A) features. See **SI 4.3**, **Figure S10** for identical plots with input λ = 1550 nm.

We experimentally characterized SHG from the MSP using pulsed input sources in a commercial laser scanning microscope, **Figure 4a (SI 6.1)**, as well as a home-built setup, **Figure 4b**. At identical input powers and detection settings, we are unable to measure the SHG from independently synthesized colloidal ZnO NRs (**SI 6.2**). This indicates an enhancement in the SHG from the MSP compared to the ZnO NR building blocks.

While the SHG from individual ZnO NRs were below our minimum detectable signal, we used full-wave FDTD simulations to estimate the enhancement in the normalized SHG conversion efficiencies η ($W^{-1}$) of different configurations of ZnO NRs, **Figure 4c-f**. In carrying out the simulation, we simplified the optical nonlinearity of ZnO NR to have effective 2$^{nd}$ order susceptibility ($\chi^{(2)}$ = 15 pm/V), whose value was within the ranges of previously reported values[38] (**SI 5.2**). When placing a single $\chi^{(2)}$ ZnO NR at the shadow side of a SiO$_2$ core μ-sphere, the photonic nanojet enhances its η by an approximate 7-fold, **Figure 4d.** In the presence of the meta-shell featuring the graded index and spherical anisotropy profile but without the $\chi^{(2)}$, there is an additional 66-fold increase in η, **Figure 4e**. Finally, when $\chi^{(2)}$ is assigned to the meta-shell in its entirety, there is an additional 87-fold increase in η, **Figure 4f**. Hence, there is an approximate $10^4$-fold enhancement in the SHG $\eta$ between an MSP2.2 and a single ZnO NR (**SI 6.3**). It is also worth noting that the spherical symmetry of the meta-shell particle makes it insensitive to polarization variations of a linearly polarized input.

We quantified the experimental value of the $\eta$ ($W^{-1}$) by the MSP2.2 utilizing custom-built NLO microscopy (**SI 7.1 – 7.6**). Taking into account the SHG signal collected in the reflection mode, $\eta$ averages to $9.97 \times 10^{-12}$ $W^{-1}$. The full-wave FDTD simulations replicating the experimental setup (**SI 7.5**) resulted in $\eta = 8.53 \times 10^{-12}$ $W^{-1}$ which closely approximates the experimental measurements.

Taking into account the SHG signal collected in the forward-scattered mode, the maximum $\eta$ reached $8.05 \times 10^{-11}$ $W^{-1}$. This corresponds to $1.08 \times 10^{-7}$ of absolute conversion efficiency, which is the highest achieved by solution processed dielectric colloidal particles reported thus far, displaying five orders of magnitude improvement compared to that reported from the BaTiO3 nanoparticles[24], and also displays two orders of magnitude improvement compared that observed with a colloidal hybrid plasmonic superstructure[39]. It should be noted that the maximum absolute conversion efficiency was achieved with the maximum available power at the particles, which was far from the damage threshold and any saturation in the efficiency, Figure 5b. While the average input power of femtosecond pulses at the particles, $P_{in,avg}$, for the above measurement was 46mW, with a different source at 1.5 um, we measured that the MSP2.2 could withstand $P_{in,avg}$ up to 1 W. The SHG measurement from both the reflected and the forward-scattered detection scheme exhibit quadratic dependence to the input power, **Figure 4b**.

The $\eta$ measurement from the forward-scattered mode averages to $\eta = 6.43 \times 10^{-11}$ $W^{-1}$ ($N = 5$), indicating a forward to backward scattering ratio (*F/B*) of 6.45 with the current measurement setup. The high *F/B* accords with the findings from the FDTD simulations. Farfield directivity polar plot, **Figure 4i**, derived by taking the FFT of the second-harmonic images, **Figure 4g**, shows enhanced forward scattering of SHG (See **SI 7.6** for farfield pattern of reflected SHG). The FDTD full-wave simulation of MSP2.2 with identical experimental parameters yields $\eta = 6.81 \times 10^{-11}$ $W^{-1}$ for the forward-scattered SHG and the *F/B*

= 7.98. Furthermore, taking into account the total SHG in all directions, the MSP2.2 in the simulation exhibits $\eta = 1.48 \times 10^{-10}\ W^{-1}$. Consistency among our 3D full-wave simulations, analytical studies, and experimental results illustrates the efficacy of our developed techniques for future studies in the colloidal nonlinear nanophotonics.

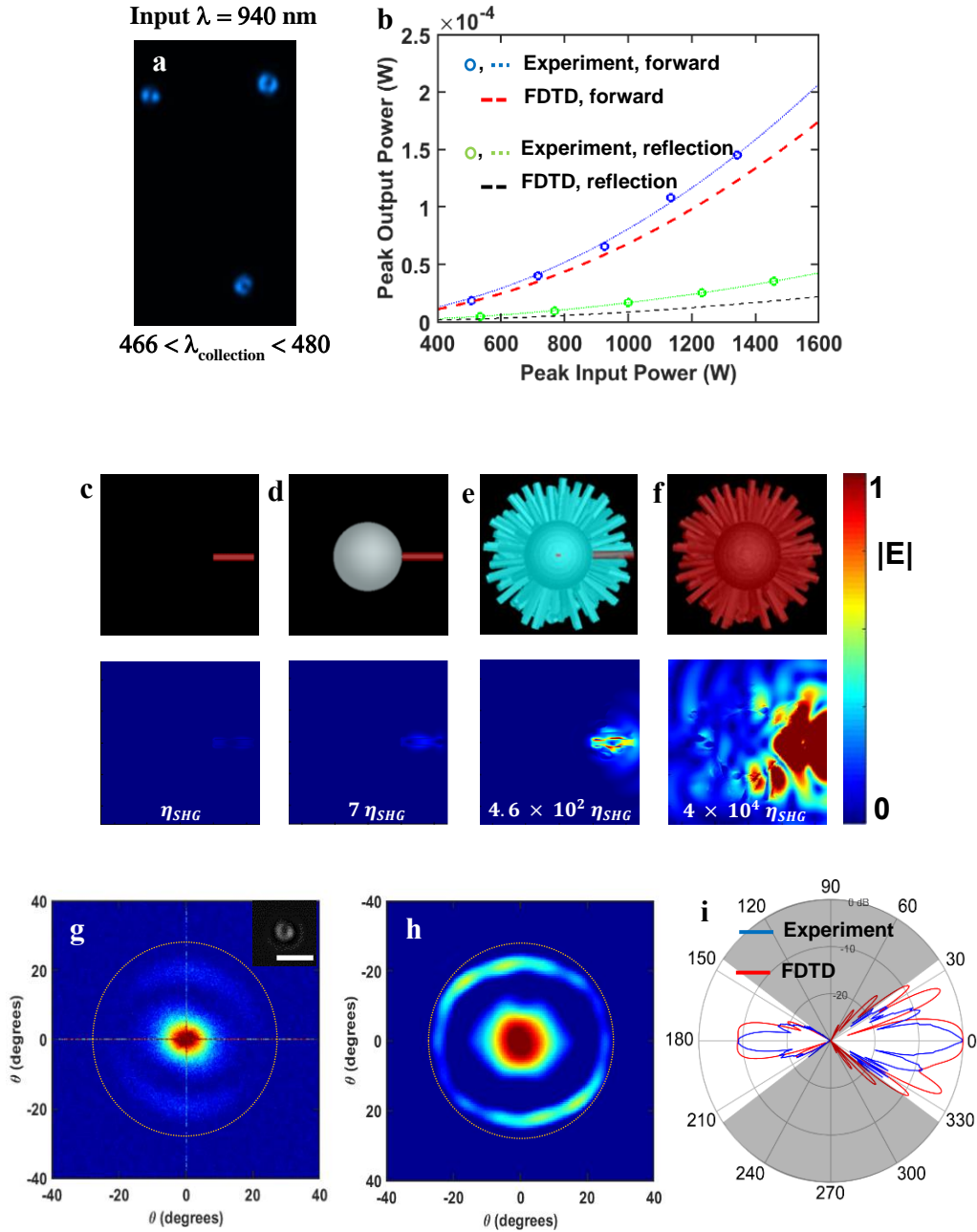

**Figure 4. Second harmonic generation of light by the MSP.** (**a**) confocal microscopy image of the SHG by the MSP2.2 (see **SI 6.1** details); (**b**) Experimental values of forward scattered and reflected SHG output power, measured from custom built optical nonlinear microscope (**SI 7.1-7.3**), display quadratic dependence to the input power. For SHG signal collected in the reflection mode, $\eta$ ranged between $6.17 \times 10^{-12}$ $W^{-1}$ and $1.53 \times 10^{-11}$ $W^{-1}$ ($N = 29$). For the SHG signal collected in the forward scattered mode, $\eta$ ranged between $4.87 \times 10^{-11}$ $W^{-1}$ and $8.05 \times 10^{-11}$ $W^{-1}$ ($N = 5$). Sample having the highest SHG conversion values collected in the reflection mode (○) and in the forward scattered mode (○) are plotted in the graph as demonstration. The dotted green line (······, $R = 0.9997$) and the dotted blue line (······, $R = 0.996$) are fit to the quadratic curves ($y = \eta x^2$) for reflected and forward scattered SHG, respectively. The FDTD simulation of the predicted quadratic relationships for the reflected (--·, $\eta = 8.53 \times 10^{-12}$ $W^{-1}$) and forward scattered (--·, $\eta = 6.81 \times 10^{-11}$ $W^{-1}$) SHG are also plotted; **c-f**, FDTD simulation, for input pulse centered at $\lambda = 1550$ nm, showing 4 orders of magnitude enhancement in the SHG conversion efficiencies $\eta$ between a model MSP2.2 and a ZnO NR; Farfield radiation pattern of the forward scattered SHG from (**g**) the experiment (the inset is the experimental SHG image) and from the (**h**) FDTD simulation. Dotted circle represents the angle of collection for the forward scattered SHG; (**i**) The directivity plot showing enhanced forward scatter of the SHG (—— experiment, —— simulation). The MSP employed for the study here are samples from **Figure 1d** (MSP2.2); All scale bar: 5μm

The presented meta-shell approach for enhancing nonlinear optical processes in nanostructures is in sharp contrast with the ongoing nonlinear nanophotonic efforts. While most of the efforts utilize high Q-factor Mie resonances confined within the structures usually comprised of high-index materials, our meta-shell approach utilizes collective interferences of high-density low-Q-factor modes in low-index materials, which lead to formation of strong hotspots. Furthermore, the metamaterial shell allows for engineering the Mie resonances of the suprapaticle to maximize the spatial overlap of the hotspot with the regions having highest density of the $\chi^{(2)}$ colloidal elements. This novel platform utilizes the low-index wide-bandgap colloidal supraparticles made of ZnO for nonlinear optical wavelength conversion. While ZnO does not have a large $\chi^{(2)}$ compared to other commonly used nonlinear materials, its well-developed chemical synthesis in the colloidal form makes it compatible with a wide array of scientific and industrial processes. Combing unique chemical and optical properties that arises from the MSP packaged into a single supraparticle could further expand development of the colloidal nonlinear nanophotonics to chemistry and biology.

**Methods**

**Chemical synthesis of MSP:** The colloidal meta-shell supraparticles (MSP) are constructed via aqueous colloidal synthesis under mild synthetic conditions. Initially, positively charged ZnO nanoparticles (NPs) (Sigma Aldrich) are electrostatically adsorbed onto negatively charged colloidal core spheres. The core spheres utilized for the construction of the MSP are polystyrene (PS) and silica ($SiO_2$) μ-spheres. We utilized carboxyl functionalized PS μ-spheres (Polysciences Inc.) that is imparted with interfacial negative charges. In order to impart interfacial negative charges to the $SiO_2$ μ-spheres, we overlaid the particle with polyelectrolyte Poly(allylamine hydrochloride) - Poly(acrylic acid) coating sequence in a layer-by-layer approach. The interfacial coating of ZnO NPs function as seeds from which ZnO nano-rods (NR) are grown in vertical orientations via a combination of hydrothermal and sonochemical processes. Synthetic methods to impart interfacial negative charges to the $SiO_2$ μ-spheres will be described in detail in the forthcoming publication "Photocatalytic Hedgehog Particles for High Ionic Strength Environment"[40]. In a typical synthesis that constructs the meta-shell of the MSP2.2 (1 μm in core dimeter, 600 nm in meta-shell thickness, total MSP diameter of 2.2 μm), the ZnO NP coated core spheres are immersed in ZnO precursors, comprised of 25 mM of zinc nitrate hexahydrate ($Zn(NO_3)_2 \cdot 6H_2O$, Sigma Aldrich) and 25 mM hexamethylenetetramine ($C_6H_{12}N_4$, Sigma Aldrich), dissolved in aqueous solution. ZnO NRs are grown by subjecting the mixture to hydrothermal (90º) and sonochemical energy for 90 minutes. The length, thickness and the densities of the ZnO NRs in the meta-shell are easily tailored by adjusting their growth conditions, such as the growth time, ZnO pre-cursor concentrations and the seeding density. The $SiO_2$ μ-spheres are synthesized by the well-known Stober process. The synthesized MSP are maintained as a dispersion in aqueous environment.

**FDTD simulation:** Full wave FDTD simulations are employed to characterize optical extinction properties of the MSP. First, we created a model MSP composed of either a $SiO_2$ or a PS core of diameter $d = 1$ μm and the meta-shell consisting a spherical array of ZnO NRs (length $l = 600$ nm, thickness $w = 120$ nm) that mirrors the experimental construct. Computer aided design software was used to reconstruct a model MSP with imperfect orthogonalization of ZnO NRs to mirror the experimental construct and to remove artifacts due to symmetry that is not present in the experimental construct (see **SI 2.2** and **SI 2.4**). TFSF source was utilized to obtain extinction cross section of the MSP.

FDTD simulations are also employed to estimate the second harmonic conversion efficiencies of the MSP. The simulation setup is configured to emulate the experimental conditions such as the objective *NA* and the pulsed input sources (see **SI 6** and **SI 7**). In running the full wave FDTD, calculations for second harmonic processes, the collection wavelengths $\lambda_{2\omega}$ is set to the second harmonic ranges and the simulation is carried out in a "nonorm" state. This removes numerical artifacts associated with the continuous wave approximation/normalization from the input pulse, $\lambda_\omega$, whose pulse length $\tau_\omega$ does not expand to the collection wavelength $\lambda_{2\omega}$. The output power, $\widetilde{P_{2\omega}}$, collected from the frequency domain field and power monitor is separately renormalized with the output pulse length $\tau_{2\omega}$, to obtain $P_{2\omega}$. The input power, $\widetilde{P_\omega}$, collected from the frequency domain monitor is renormalized with the pulse length $\tau_\omega$. This represents the peak power, $P_\omega$, delivered to the model MSP.

$$P_{2\omega} = \frac{\widetilde{P_{2\omega}}}{\tau_{2\omega}^2}$$

$$P_\omega = \frac{\widetilde{P_\omega}}{\tau_\omega^2}$$

The normalized SHG conversion efficiency, $\eta_{SHG}$, is obtained by dividing the output power by the input power squared.

$$\eta_{SHG} = \frac{P_{2\omega}}{P_\omega^2} = \frac{\widetilde{P_{2\omega}}}{\widetilde{P_\omega}^2} \times \frac{\tau_\omega^4}{\tau_{2\omega}^2}$$

In calculating the conversion efficiency, we simplified the optical nonlinearities of ZnO NR to have isotropic susceptibility tensor having an effective $\chi^{(2)} = 15$ pm/V. The *reference 41* reports a range of nonlinear coefficients from $d_{eff}$ = 2 pm/V to 15 pm/V [38,41]. We have chosen a median value ($d_{eff}$ = 7.5 pm/V) from the reported range to numerically approximate second harmonic conversion of the MSP.

**NLO microscopy for SHG measurement:** The schematic in **Figure S20 (SI 7.1)** depicts the nonlinear optical (NLO) microscopy setup constructed to detect the SHG generated by the MSP. **Table S1 (SI 7.1)** shows the optical components employed in the setup. Ultra-short pulse frequency comb (Menlo Systems, $\tau$ = 80 fs, $f_{rep}$ = 250 MHz) centered at $\lambda_{in}$ = 1550 nm is used as the input source. The input pulse is subject to a continuously variable neutral density filter wheel to obtain variable input average power. The input beam is guided into a 50x objective (Mitutoyo Plan APO NIR, $NA$ = 0.42) and focused onto a single particle on the sample slide with a spot size $d_f \approx$ 2.6 μm. A collimated LED white light source was incorporated to locate and focus the input pulse into individual MSP. The backscattered SHG is subject to a pair of filters to remove the input light and the third harmonic generation. A 1" VIS lens was used to focus the SHG onto a photodiode power sensor. The total average power of the backscattered SHG is then determined by accounting for the power loss through each optical component. The transmission values of each optical component that the SHG passes through are shown in **Table S2**.

The forward scattered SHG was collected via aspheric condenser lens *(NA .79, F = 16 mm)* and subject to a pair of filters to remove the input pump and the higher order harmonic generations identical to the reflection mode setup. The total average power of the forward scattered SHG is then determined by accounting for the power loss through each optical component, listed in **Table S2**. The aspheric condenser lens was replaced with a 60X objective (Nikon, M Plan 60, 630770) when taking the forward scattered images to obtain the farfield pattern.

**Confocal microscopy:** The SHG measurement was also carried out with confocal microscopy in which the MSP are irradiated with femtosecond pulses (140fs, 80MHz) centered at $\lambda$ = 900 nm and at 5% of its maximum available power. In order to fully immerse the MSP within the spotsize, a low *NA* (0.16) 5X objective was use. The SHG intensities obtained from the images are subtracted by the background noise, normalized by the gain settings, followed by normalization with the square of the input power reading.

**Analytical modeling:** The average unit-cell size of the nanowires at the outer interface of the particle is sub-wavelength. Hence, we can approximate the nanowires using an effective medium approach. Since the nanowires are mostly oriented radially, we can model them with all-dielectric radial anisotropy using Maxwell-Garnett approximation. The effective permittivity parallel and normal to the direction of the nanowires are[42]:

$$\varepsilon_\perp = \rho(r)\varepsilon_d + (1-\rho(r))\varepsilon_h,$$

$$\varepsilon_\| = \frac{(1+\rho(r))\varepsilon_d\varepsilon_h + (1-\rho(r))\varepsilon_h^2}{(1-\rho(r))\varepsilon_d + (1+\rho(r))\varepsilon_h}$$

where $\varepsilon_d$ is the permittivity of the dielectric nanowire, $\varepsilon_h$ is the permittivity of the host medium (which is air here), and $\rho(r)$ is the nanowire filling factor. Since the density of the nanowires reduces as the distance from the center increases, the filling factor is *r*-dependent. This results in a graded-index profile in the shell. For the TE (magnetic) modes, the electric field in the *r* direction is zero. Hence, the TE modes do not feel the anisotropy of the shell. However, the TM (electric) modes are affected by the anisotropy of the shell.

To calculate the scattered fields, we have used the Mie theory. We have assumed the input is a plane wave polarized in the *x* direction and propagates in the *z* direction. The input electric field intensity is $E_0$. The electric and magnetic fields of TE and TM modes in the *r* direction in the m[th]-medium can be written as[36]:

$$E_r^m(r,\theta,\varphi) = A_m \frac{E_0}{k_0^2 r^2}\cos\varphi \sum_{n=1} i^{(n+1)}(2n+1)\, z_{n_{e,m}}\!\left(k_0\sqrt{\varepsilon_{\perp,m}}\,r\right) P_n^{(1)}(\cos\theta),$$

$$H_r^m(r,\theta,\varphi) = B_m \frac{E_0}{\eta k_0^2 r^2} \cos\varphi \sum_{n=1} i^{(n+1)}(2n+1) z_n\left(k_0\sqrt{\varepsilon_{\perp,m}}r\right) P_n^{(1)}(\cos\theta),$$

where $z_n(x)$ is a spherical Bessel function of $n^{th}$-order, $P_n^{(1)}$ is associate Legendre function of $1^{st}$ order and $n^{th}$ degree, $\varepsilon_{\perp,m}$ is the effective transverse permittivity of the $m^{th}$-medium, and $\eta$ is the impedance of the frees-pace. The order of the Bessel functions in anisotropic media is:

$$n_{e,m} = \sqrt{\frac{\varepsilon_{\perp,m}}{\varepsilon_{\parallel,m}} n(n+1) + \frac{1}{4} - \frac{1}{2}}.$$

It is seen that by changing the anisotropy, we can control the index of the Bessel function for the electric modes.

To model the graded-index profile, we have discretized the metamaterial shell to 40 homogenous layers. By applying the boundary conditions at the interface between each layer, we can find the scattering coefficients of the particle with metamaterial shell.

AUTHOR INFORMATION

**Corresponding Author**

Alireza Marandi (marandi@caltech.edu)

**Present Addresses**

†If an author's address is different than the one given in the affiliation line, this information may be included here.

**Author Contributions**

The manuscript was written through contributions of all authors. All authors have given approval to the final version of the manuscript. ‡These authors contributed equally

**References**

At the end of the document

# Supplementary Information

# Table of Contents

1. **Experiment – construction of meta-shell supraparticles (MSP)**

**1.1 Chemical synthesis of MSP**

**Comment 1:** The colloidal meta-shell supraparticles (MSP) are constructed via aqueous colloidal synthesis under mild synthetic conditions. Initially, positively charged ZnO nanoparticles (NPs) (Sigma Aldrich) are electrostatically adsorbed onto negatively charged colloidal core spheres. The core spheres utilized for the construction of the MSP are polystyrene (PS) and silica ($SiO_2$) µ-spheres. We utilized carboxyl functionalized PS µ-spheres (Polysciences Inc.) that is imparted with interfacial negative charges. In order to impart interfacial negative charges to the $SiO_2$ µ-spheres, we overlaid the particle with polyelectrolyte Poly(allylamine hydrochloride) - Poly(acrylic acid) coating sequence in a layer-by-layer approach. The interfacial coating of ZnO NPs function as seeds from which ZnO nano-rods (NR) are grown in vertical orientations via a combination of hydrothermal and sonochemical processes. Synthetic methods to impart interfacial negative charges to the $SiO_2$ µ-spheres will be described in detail in the forthcoming publication "Photocatalytic Hedgehog Particles for High Ionic Strength Environment"[1]. In a typical synthesis that constructs the meta-shell of the MSP2.2 (1 µm in core dimeter, 600 nm in meta-shell thickness, total MSP diameter of 2.2 µm) , the ZnO NP coated core spheres are immersed in ZnO precursors, comprised of 25 mM of zinc nitrate hexahydrate ($Zn(NO_3)_2 \cdot 6H_2O$, Sigma Aldrich) and 25 mM hexamethylenetetramine ($C_6H_{12}N_4$, Sigma Aldrich), dissolved in aqueous solution. ZnO NRs are grown by subjecting the mixture to hydrothermal (90º) and sonochemical energy for 90 minutes. The length, thickness and the densities of the ZnO NRs in the meta-shell are easily tailored by adjusting their growth conditions, such as the growth time, ZnO pre-cursor concentrations and the seeding density. The

SiO$_2$ µ-spheres are synthesized by the well-known Stober process. The synthesized MSP are maintained as a dispersion in aqueous environment.

**Comment 2:** There are no synthetic restrictions to the choice of the core materials in constructing the MSP due to relaxation in the lattice-matching requirements for vertical orientation of 1D ZnO on any substrates[2–4]. The core sphere is an important design parameter as dielectric µ-sphere itself holds a wealth of EM modes[5,6]. Of particular interest is their collective interferences that lead to formation of a local $|E|$ hotspot at the shadow side of the particle upon light incidence, known as the photonic nanojet[7,8]. For an efficient SHG conversion, we want to maximize the spatial overlap of the $|E|$ hotspot with the interfacial array of $\chi^{(2)}$ ZnO NRs that constitutes the meta-shell. This requires the core refractive index to be $n < 2$ [7]. Despite a polystyrene (PS) core best accomplishing the criteria (**SI 3.2, 3.3**), we focused our studies with the MSP having a SiO$_2$ core.

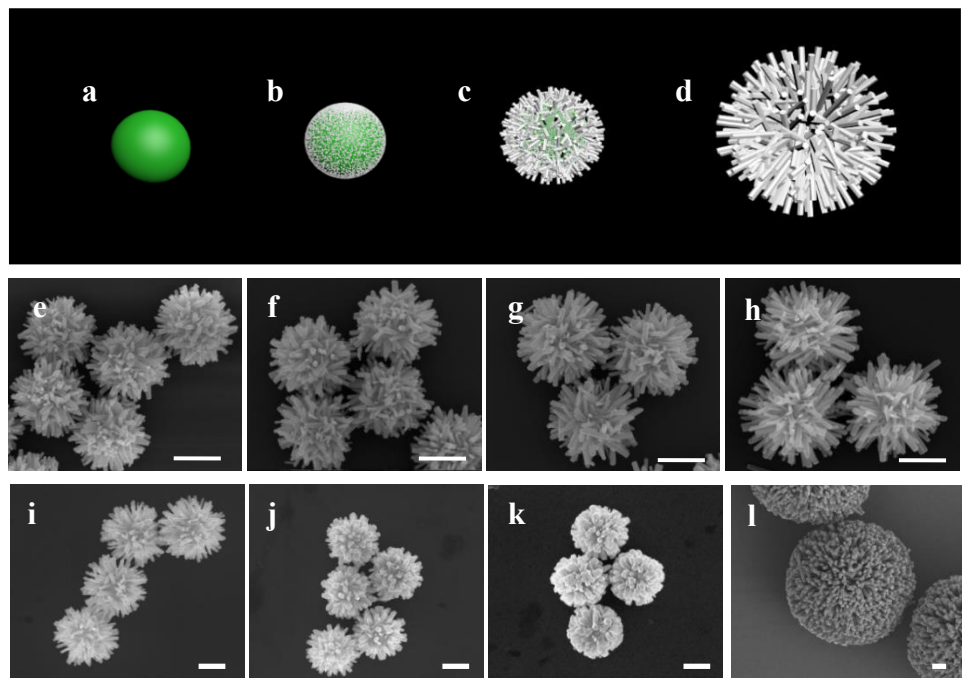

**Figure S1. Meta-shell supraparticles. a-d,** The MSP are synthesized by adsorbing positively charged ZnO NPs onto negatively charged dielectric μ-sphere, from which ZnO NRs are grown orthogonal to the sphere interface via a combination of hydrothermal and sonochemical processes; **e-h**, SEM images of the MSP with core particle of $d = 1$ μm and ZnO NRs of different lengths, **(e)** $l \approx 0.19$ μm, **(f)** $l \approx 0.27$ μm, **(g)** $l \approx 0.4$ μm, **(h)** $l \approx 0.6$ μm; **i-k**, SEM images of MSP with core particle of $d = 1$ μm and ZnO nanospikes of different widths, **(e)** $w \approx 0.19$ μm, **(f)** $w \approx 0.27$ μm, **(g)** $w \approx 0.4$ μm; **(l)** MSP with core particle $d = 6$ μm; scale bar = 1 μm

## 1.2 UV-Vis spectroscopy – extinction spectra of MSP

**Comment:** We first investigated light extinction spectra of the MSP as this provides us opportunity for an initial estimation and prediction of possible EM modes and their spectral locations supported by the supraparticle. The first series of the MSP were initially constructed with mono-disperse carboxyl functionalized polystyrene (PS) core spheres due to their commercial availability and due to simpler synthetic protocols in constructing the MSP. For studying optical nonlinear conversion by the MSP, the PS cores were subsequently replaced with $SiO_2$ core owing to its superior mechanical and thermal stability against high power laser. We first estimated the extinction characteristics of the MSP in their aqueous colloidal dispersion state. Two extinction peaks, atypical of low-index dielectric colloid for its representative size ranges, are observed. They are one broadband peak at ~ 730 nm ($P_1$) and another narrower peak at ~405nm ($P_2$), **Figure 1f** in the main text. For obtaining light extinction spectra in the aerosol format, lyophilized powder of MSP2.2 are placed in a quartz cuvette and suspended in mid-air via introduction of pressurized $N_2$ gas.

## 2. FDTD Simulations - extinction cross-section

### 2.1 Extinction cross-section of ZnO NRs and PS µ-spheres in aqueous dispersion

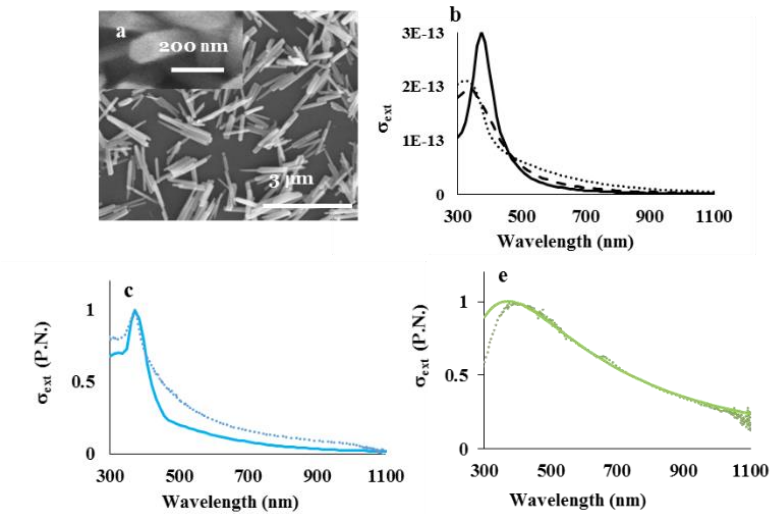

**Figure S2. Extinction spectra of ZnO NR and PS µ-sphere, comparison between the spectroscopy and the FDTD simulation, in aqueous dispersion.** (**a**) Scanning electron microscopy (SEM) image of independently synthesized ZnO NRs. The inset is a magnified view of the ZnO NRs in the meta-shell constructed on a core sphere that shows hexagonal cross-section. (**b**) FDTD simulation showing the extinction cross section ($\sigma_{ext}$) of a single ZnO NR ($w = 120$nm, $l = 600$nm) that is parallel to (solid line), at $45°$ angle to (dotted line) and perpendicular to (perforated line) the incident light; (**c**) FDTD simulation showing peak normalized extinction cross section ($\sigma_{ext}$, P.N.) of a single ZnO NR where the maximum extinction value per wavelength point are taken between the 3 incident angles (solid) vs. experimental the P.N. extinction of ZnO NRs measured with UV-Vis spectroscopy (dotted); (**d**) FDTD simulation showing the P.N $\sigma_{ext}$ of a PS µ-sphere (solid), $d = 1$ µm, vs. the P.N. experimental extinction value from the spectroscopic measurement (dotted).

**Comment:** Lumerical FDTD Solutions are employed to characterize the optical properties of the colloidal particles. The finite difference time domain (FDTD) calculations of the extinction cross-section ($\sigma_{ext}$) of PS µ–spheres illuminated with total field scattered field (TFSF) plane wave source approximates the spectral line shape of the optical extinction from the spectroscopic measurement with good agreement. The FDTD simulations of the $\sigma_{ext}$ of ZnO NRs also shows good agreement with the extinction measurement. A model ZnO NR is constructed to have length $l = 600$ nm, width $w = 120$ nm and a hexagonal cross-section. Larger discrepancy is due to polydispersity in the synthesized ZnO NRs.

## 2.2 Model MSP and its extinction cross-section in aqueous dispersion

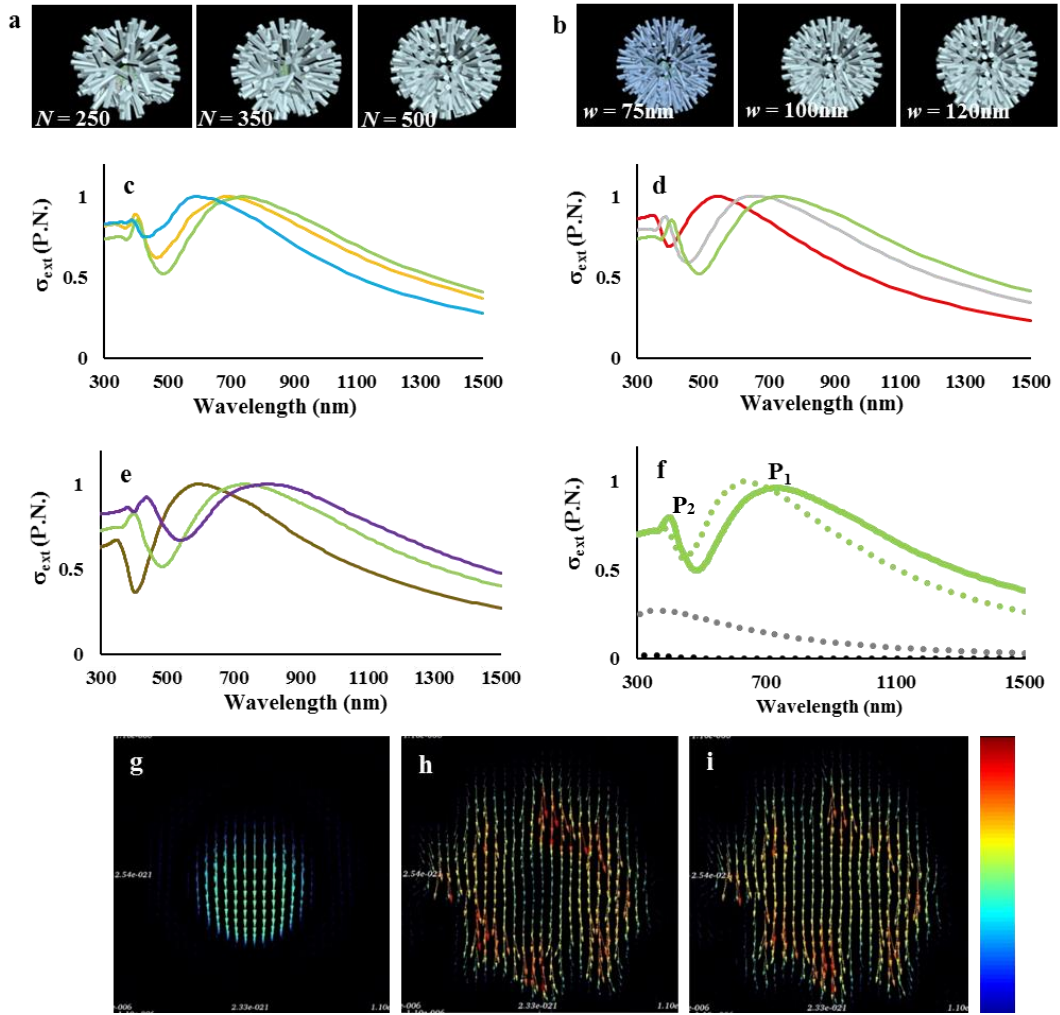

**Figure S3. Model MSP and FDTD simulations. a-b,** Model MSP is constructed, via computer aided design (CAD) software, by layering ZnO nanoparticles (pyramids, 15 nm) on $SiO_2$ or PS μ-sphere, onto which vertically oriented ZnO NRs are scattered to create imperfect orthogonalization that reflect the experimental construct. The model MSP are imported into the FDTD simulations; the CAD of model MSP with meta-shell having varying ZnO NR (**a**) densities and (**b**) thicknesses. **c-f**, FDTD simulations showing the P.N. $\sigma_{ex}$ of the MSP having (**c**) variations in the ZnO NR densities $N$ (constant $l = 600$ nm and $w = 120$ nm) with $N = 500$ (green), $N = 350$ (orange), $N = 250$ (blue) and (**d**) variations in the ZnO NR width $w$ (constant $l = 600$ nm and $N = 500$) with $w = 120$ nm (green), $w = 100$ nm (grey), $w = 75$ nm (red) and (**e**) variations in the ZnO NR length $l$ (constant $w = 120$ nm and $N = 500$) with $l = 300$ nm (brown) where $l = 600$ nm (green), $l = 900$ nm (purple); (**f**) evolution of $\sigma_{ext}$ starting from single ZnO NR (black, dotted), PS μ-sphere (grey, dotted) to meta-shell only (spherical array of ZnO NRs, $N = 500$, $l = 120$ nm, $w = 120$ nm, green, dotted) to the MSP (green, solid); **g-i**, FDTD simulations, at incident $\lambda = 730$ nm, displaying electric field profile at the center cross-section of a (**g**) PS μ-sphere ($d = 1$ μm), (**h**) meta-shell only and (**i**) the MSP. All FDTD simulations carried out here is modeled as aqueous dispersion of MSP to emulate UV-Vis spectroscopic measurement procedure.

**Comment:** FDTD simulations are employed to characterize the optical linear properties of the MSP. We created a model MSP composed of either a $SiO_2$ or a PS core of diameter $d = 1$ μm and the meta-shell consisting a spherical array of ZnO NRs (length $l = 600$ nm, thickness $w = 120$ nm) that mirrors the experimental construct. Computer aided design software was used to reconstruct a model MSP with imperfect orthogonalization of ZnO NRs to mirror the experimental construct and to remove artifacts due to symmetry that is not present in the experimental construct (**Figure S3a, b**). The FDTD full wave simulations of the $\sigma_{ext}$ by the model MSP2.2 subject to TFSF source approximates the experimental extinction line shape from the UV-Vis spectroscopy measurement with good agreement, **Figure 1j** in the main text. The simulation also recaptures the ZnO NR density-, width-, and length-dependent spectral redshift in the extinction peaks, **Figure S3c-e**.

The lack of spectral correlation between the MSP and its components, the ZnO NR and the core sphere (**Figure S2**), suggest that the peak $P_1$ originate from the supraparticle architecture. Indeed, the evolution of the $\sigma_{ext}$ from a ZnO NR, PS μ-sphere, meta-shell only and to the model MSP shows that the ortho-spherical arrangement of the ZnO NR array, that constitutes the meta-shell, is the governing structural attribute that generates the broadband peak, **Figure S3f**. The broadband peak further redshifts to $P_1$ in the presence of the core sphere. This is due to increase in the electric polarizability that extends the electric field profile through the core substrate, **Figure S3g-i**.

## 2.3 Simplified triple-shell core model and its extinction cross-section in aqueous dispersion

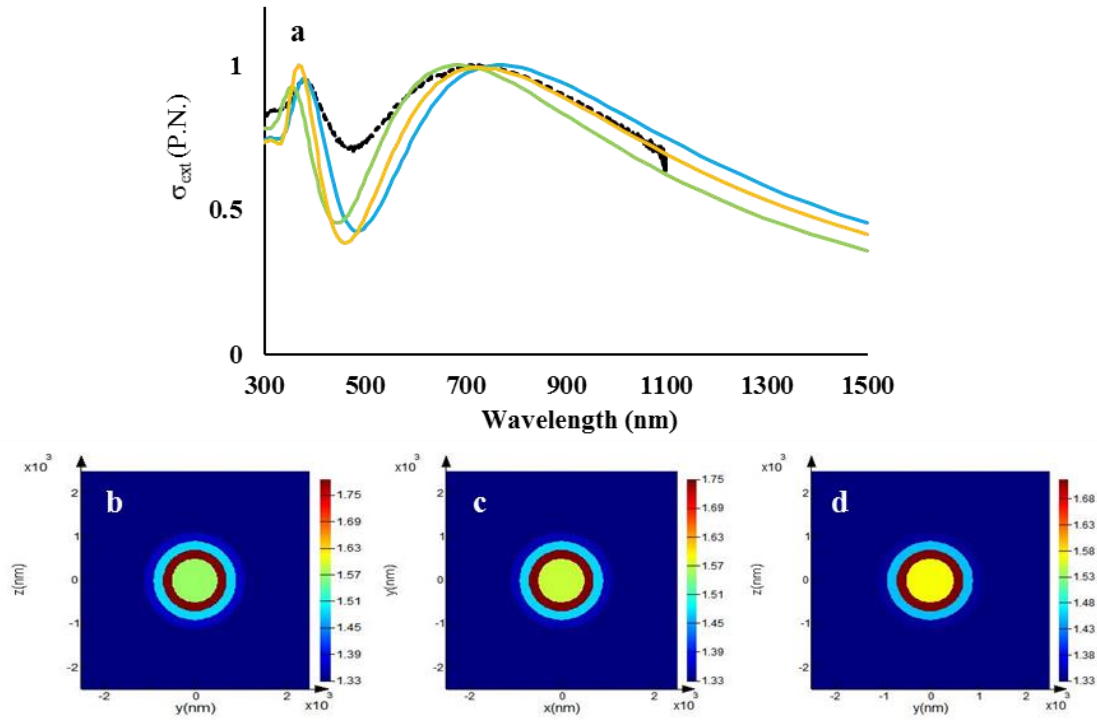

**Figure S4. Discretized gradient refractive index triple-shell model;** triple-shell model having 1:1:1 thickness ratios with graded variations in the refractive indices, where the refractive index of each shell layers are determined by Maxwell Garnett effective medium approximations; (**a**) FDTD simulations of the P.N. $\sigma_{ext}$ of the triple-shell model that corresponds to MSP having ZnO NR of $w = 120$ nm, $l = 600$ nm and $N = 275$ (**blue**), $N = 265$ (**orange**), $N = 250$ (**green**). The perforated black line represents the spectroscopic measurement (A.U.) of MSP in aqueous suspension. Color map of the refractive indices in the 1:1:1 triple-shell that corresponds to (**b**) $N = 275$ ($f_{outer} = 0.27$, $f_{middle} = 0.42$, $f_{inner} = 0.75$), (**c**) $N = 275$ ($f_{outer} = 0.26$, $f_{middle} = 0.41$, $f_{inner} = 0.72$), (**d**) $N = 250$ ($f_{outer} = 0.25$, $f_{middle} = 0.39$, $f_{inner} = 0.68$), where $f$ is the volume fraction of ZnO.

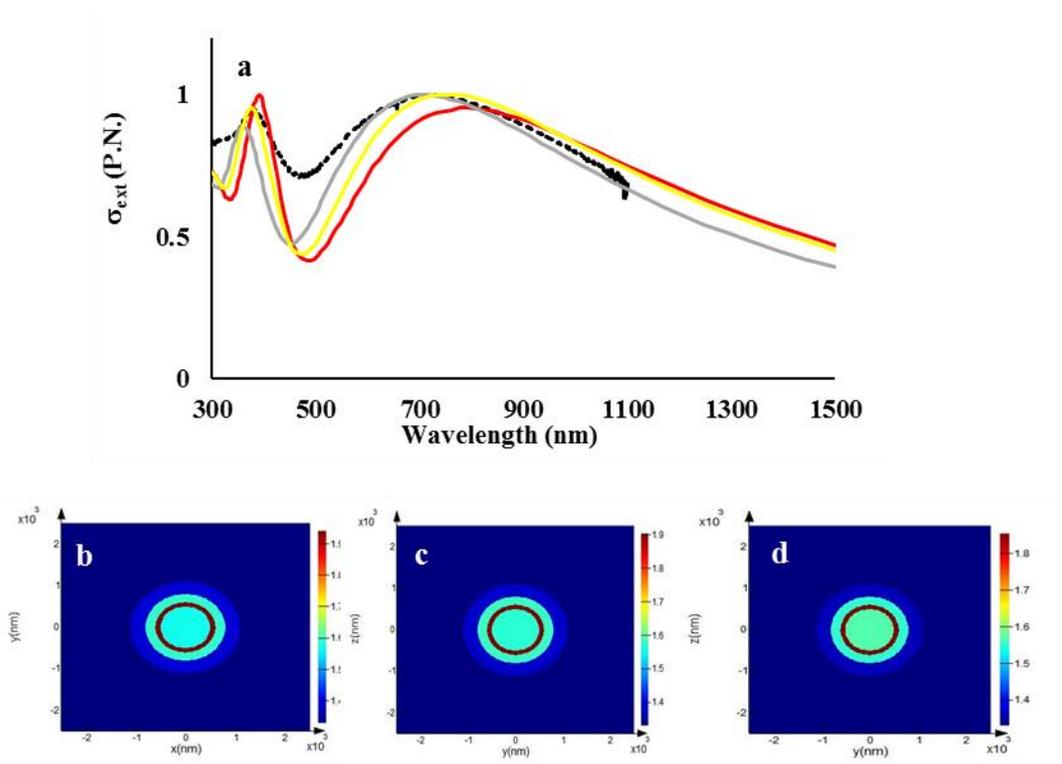

**Figure S5. Discretized gradient refractive index triple-shell model.** triple-shell model having 3:2:1 (outer:middle:inner) thickness ratios with graded variations in the refractive indices, where the refractive index of each shell layers are determined by Maxwell Garnett effective medium theory; (**a**) FDTD simulations of the P.N. $\sigma_{ext}$ of the triple-shell model that corresponds to the MSP having ZnO NR of $w =$ 120 m, $l =$ 600 nm and $N =$ 275 (**red**), $N =$ 265 (**yellow**), $N =$ 250 (**grey**). The perforated black line represents the spectroscopic measurement of the MSP in aqueous suspension. . Color map of the refractive indices in the 3:2:1 triple-shell that corresponds to (**b**) $N =$ 275 ($f_{outer} = 0.3$, $f_{middle} = 0.55$, $f_{inner} = 0.90$), (**c**) $N =$ 265 ($f_{outer} = 0.29$, $f_{middle} = 0.53$, $f_{inner} = 0.87$), (**d**) $N =$ 250 ($f_{outer} = 0.27$, $f_{middle} = 0.50$, $f_{inner} = 0.82$), ), where $f$ is the volume fraction of the ZnO.

**Comment:** It should be noted that the model MSP2.2 requires higher nanospike densities ($N$) to match the $\sigma_{ext}$ spectra of the FDTD calculations with the spectroscopic measurement from the experimental construct ($N_{experiment} \sim 200 - 250$, $N_{model} = 500$). This may be due to large unit mesh size applied in the calculation so as to remain within the computational budget. Such may not grasp the geometrical finesse and complexities within the meta-shell. On the other hand, simplifying the empirical meta-shell features into a simplified core-shell model should yield extinction line shape with good agreement between the simulation and the spectroscopic measurements. Due to vertical orientation of high aspect-ratio ZnO NR in a spherical format, the meta-shell is represented with radial graded refractive index profile with 3 discretized shells at 1:1:1 or 3:2:1 (outer: middle: inner shell) thickness ratios, layered on the core sphere. Refractive indices, $n$, of each shells are assigned by the effective medium approximation (EMA) using Maxwell Garnett mixing rule for two-phase heterogeneous media[9,10]. The FDTD calculations of the $\sigma_{ext}$ of the triple-shell model with $N_{3-shell} = 250 \sim 275$, which is a close representation of the experimental MSP2.2, indeed shows good agreement with the experimental measurement, **Figure S4, S5**.

### 2.4 Model MSP and its extinction cross-section in air

**Comment:** In carrying out the FDTD simulations to probe the optical properties of the MSP2.2 in an aqueous environment, the model MSP2.2 required ZnO NR density of $N = 500$ to achieve a good match with the spectroscopic measurements (**Section 2.2**). Interestingly, in carrying out the FDTD simulations of the $\sigma_{ext}$ with the model MSP2.2 in air, the ZnO NR density of $N = 300$ was sufficient to yield good match with the experiment. This may be due to higher visibility of the surface corrugation within the meta-shell owing to higher index contrast. The latter model MSP2.2 *(*with $N = 300$) is a closer representation to the empirical observations in the experimental construct ($N = 200 - 250$). There is a good agreement in the extinction lineshape between the FDTD simulations and the experimental spectroscopic measurement (**Figure 1k**, main text).

# 3 FDTD Simulations – Photonic nanojet in *MSP*

## 3.1 Engineering photonic nanojet with meta-shell

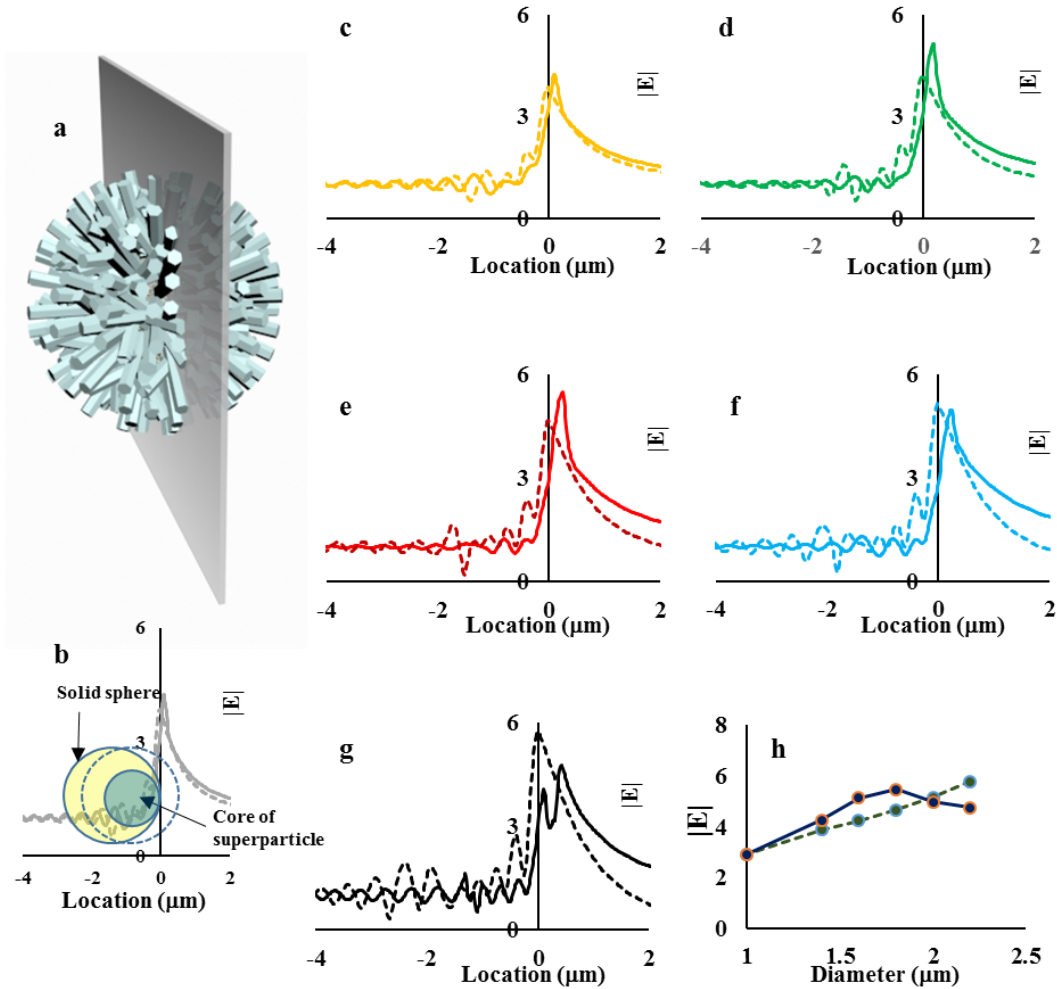

**Figure S6. Effect of meta-shell thickness on the photonic nanojet generated by the MSP.** Comparison of the electric field strength |E| in the photonic nanojet formed by the MSP and that formed by a solid sphere of equivalent diameter to the MSP. The refractive index of the solid sphere in comparison is the same as that of the core in the MSP. FDTD simulation is carried out at incident $\lambda = 900$ nm and the |E| profile is plotted along the (**a**) middle cross-section of the MSP (or solid sphere). The thickness of the MSP is varied while the $SiO_2$ core diameter is kept constant at 1 µm. (**b**) $x = 0$ coordinately coincides at the interface between the meta-shell and the core in the MSP and at the solids sphere interface (yellow), both at their shadow side; **c-g**, comparison of |E| in the photonic nanojet formed by MSP (solid) and that by the solid $SiO_2$ sphere (perforated) of equivalent diameters, (**c**) $d = 1.4$ µm, (**d**) $d = 1.6$ µm, (**e**) $d = 1.8$ µm, (**f**) $d = 2$ µm, (**g**) $d = 2.2$ µm; (**h**) peak |E| along the middle cross-section of the photonic nanojets formed by the MSP (solid) and by the $SiO_2$ solid spheres of equivalent diameters (perforated).

**Comment:** In this section, we employed FDTD simulation to observe the effect of the meta-shell thickness on the photonic nanojet features generated by the MSP. The core refractive index and its diameter are kept constant ($SiO_2$, $d = 1$ µm), while the thickness of the meta-shell is varied. The photonic nanojet features observed in the MSP are also compared with that formed by a solid $SiO_2$ sphere with equivalent overall diameter to the MSP. In the graphs in **Figure S6c-g**, $x = 0$ coordinately coincides at the interface between the core and the meta-shell in the MSP at its shadow side, **Figure S6b**. The $x = 0$ also coordinately coincides at the interface of the solid $SiO_2$ sphere at its shadow side.

In **Figure S6h,** $d = 1$ µm represents the bare core sphere (MSP without meta-shell). Compared to a bare $SiO_2$ core sphere, peak electric field strength $|E|$ in the photonic nanojet is always higher in the presence of a meta-shell.

When the overall diameter $d < 2$ µm, the peak $|E|$ in the photonic nanojet is higher in the MSP than that produced by the $SiO_2$ solid sphere of equivalent diameter to the MSP, **Figure S6c-e**. When $d > 2$ µm the trend reverses, **Figure S6f, g**. This is due to scarcity in the ZnO NR density at the peripheral edges, with increasing meta-shell thickness, which weakens its effective size. It can also be seen that the spatial location of the peak $|E|$ in the photonic nanojet hotspot proceeds away from the core interface with increasing meta-shell thickness. Meanwhile, the photonic nanojet hotspot remain fixed at the interface for the $SiO_2$ solid sphere, despite variations in their sizes.

## 3.2 Stronger $|E|$ hotspot in the photonic nanojet in the presence of a meta-shell

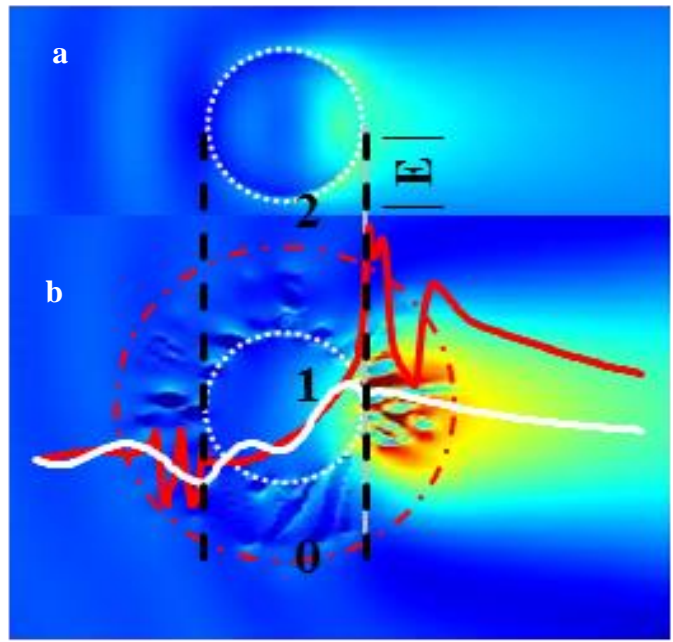

**Figure S7. FDTD simulation showing stronger $|E|$ in the photonic nanojet hotspot in the presence of a meta-shell.** Full wave FDTD simulation that compares the peak $|E|$ in the photonic nanojet hotspot generated by a (**a**) $SiO_2$ core sphere alone (white) and that by (**b**) the model MSP2.2 (red). (**a**) In the case of the bare core sphere, the region of peak $|E|$ is confined within the core sphere, (**b**) In the case of the MSP, the region of peak $|E|$ is localized in the meta-shell comprised of the $\chi^{(2)}$ nanostructures. The diameter of the core sphere is $d = 1$ μm, thickness of the meta-shell is 600 nm. The permittivity of the sphere is $\varepsilon = 2.2$. The input wavelength is $\lambda = 1550$ nm.

## 3.3 Influence of core refractive index on the photonic nanojet profile

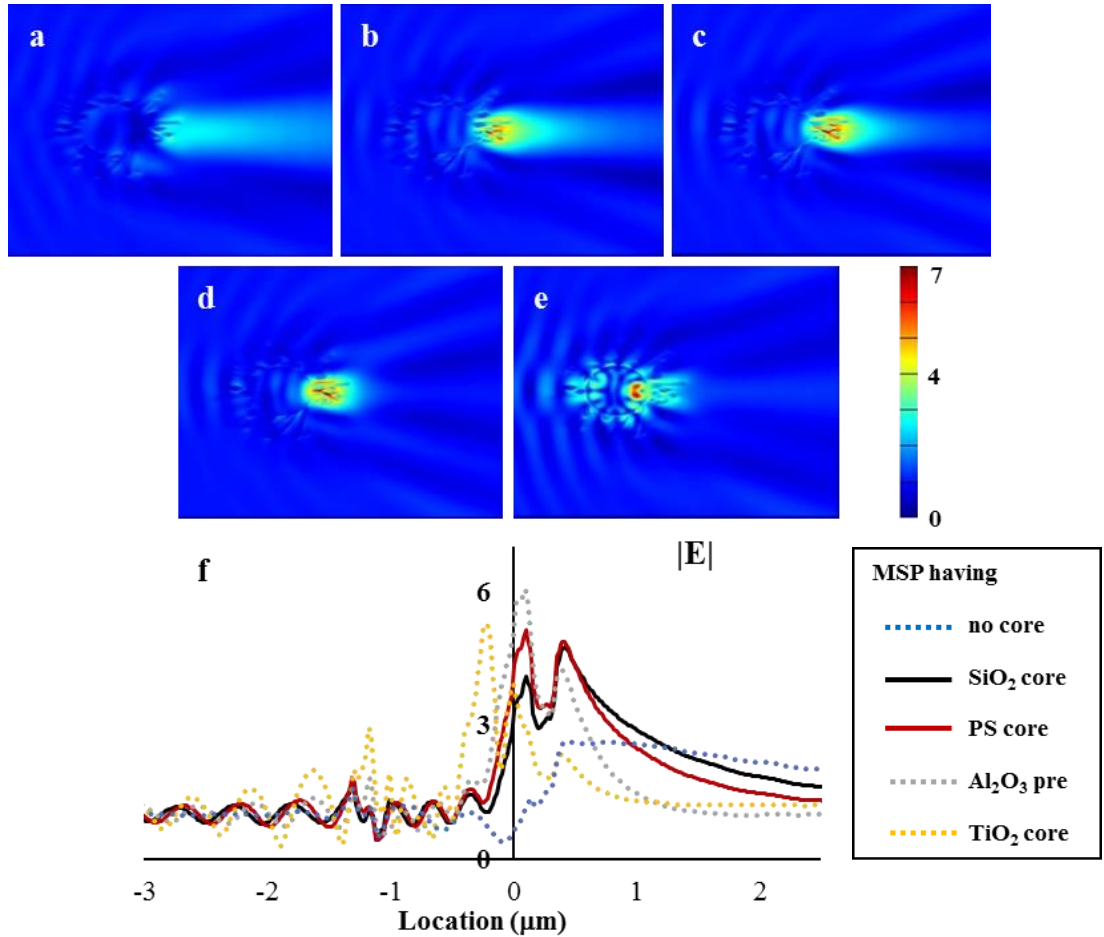

**Figure S8. Effect of the refractive index *n* of the core sphere in the MSP on the photonic nanojet.** FDTD full wave simulation displaying electric field profile at the middle cross-section of the photonic nanojet formed by the MSP having (**a**) no core and (**b**) SiO$_2$ (*n* = 1.47), (**c**) PS (*n* = 1.57), (**d**) Al2O$_3$ (*n* = 176), (**e**) TiO$_2$ (*n* = 2.5) core sphere (*d* = 1 μm). Increase in the refractive index *n* of the core sphere increases the peak |E|, but also shifts the hotspot volume towards the center of the particle. For TiO$_2$ core, the region of peak |E| is confined near the interface but concentrated within the core sphere. (**f**) Electric field strength |E| long the middle cross-section of the MSP having no core (blue dotted), SiO$_2$ core (black), PS core (red), Al$_2$O$_3$ core (grey dotted) and TiO$_2$ core (orange dotted).

**Comment:** Here, we investigated changes in the photonic nanojet features with variations in the core refractive index *n*. The geometrical features and materials properties of the meta-shell is kept constant. Increase in $n$ increases the peak electric field strength |E| in the photonic nanojet formed by MSP. This is due to higher mode densities supported by the higher index dielectric spheres whose collective interference yields higher |E|. Furthermore, increase in $n$ causes migration of the photonic nanojet hotspot towards the core interface. At sufficiently high $n$ ($\geq 2$), the photonic nanojet diminishes and the hotspots are dispersed within the core sphere. This is an unsuitable feature for optical nonlinear generations with the MSP as most of the input power does not overlap with the $\chi^{(2)}$ nanostructures. Ideally, core sphere should be chosen such that the MSP generate the hotspot whose mode volume maximizes the $\chi^{(2)}$ nanostructure coverage and whose |E| maximum intensity coincides with the spatial location of highest $\chi^{(2)}$ nanostructure densities. Despite the PS core best describing the aforementioned condition, we have chosen the $SiO_2$ core for our studies due to its higher stability against high power laser pulses.

## 4. Analytical Calculations

### 4.1 Modeling

The average unit-cell size of the nanowires at the outer interface of the particle is sub-wavelength. Hence, we can approximate the nanowires using an effective medium approach. Since the nanowires are mostly oriented radially, we can model them with all-dielectric radial anisotropy using Maxwell-Garnett approximation. The effective permittivity parallel and normal to the direction of the nanowires are[11]:

$$\varepsilon_\perp = \rho(r)\varepsilon_d + \big(1 - \rho(r)\big)\varepsilon_h,$$

$$\varepsilon_\| = \frac{(1+\rho(r))\varepsilon_d\varepsilon_h + (1-\rho(r))\varepsilon_h^2}{(1-\rho(r))\varepsilon_d + (1+\rho(r))\varepsilon_h}$$

where $\varepsilon_d$ is the permittivity of the dielectric nanowire, $\varepsilon_h$ is the permittivity of the host medium (which is air here), and $\rho(r)$ is the nanowire filling factor. Since the density of the nanowires reduces as the distance from the center increases, the filling factor is $r$-dependent. This results in a graded-index profile in the shell. For the TE (magnetic) modes, the electric field in the $r$ direction is zero. Hence, the TE modes do not feel the anisotropy of the shell. However, the TM (electric) modes are affected by the anisotropy of the shell.

To calculate the scattered fields, we have used the Mie theory. We have assumed the input is a plane wave polarized in the $x$ direction and propagates in the $z$ direction. The input electric field intensity is $E_0$. The electric and magnetic fields of TE and TM modes in the $r$ direction in the $m^{th}$-medium can be written as[12]:

$$E_r^m(r,\theta,\varphi) = A_m \frac{E_0}{k_0^2 r^2} \cos\varphi \sum_{n=1}^{\infty} i^{(n+1)}(2n+1) z_{n_{e,m}}\left(k_0\sqrt{\varepsilon_{\perp,m}}\,r\right) P_n^{(1)}(\cos\theta),$$

$$H_r^m(r,\theta,\varphi) = B_m \frac{E_0}{\eta k_0^2 r^2} \cos\varphi \sum_{n=1}^{\infty} i^{(n+1)}(2n+1) z_n\left(k_0\sqrt{\varepsilon_{\perp,m}}\,r\right) P_n^{(1)}(\cos\theta),$$

where $z_n(x)$ is a spherical Bessel function of $n^{th}$-order, $P_n^{(1)}$ is associate Legendre function of $1^{st}$ order and $n^{th}$ degree, $\varepsilon_{\perp,m}$ is the effective transverse permittivity of the $m^{th}$-medium, and $\eta$ is the impedance of the frees-pace. The order of the Bessel functions in anisotropic media is:

$$n_{e,m} = \sqrt{\frac{\varepsilon_{\perp,m}}{\varepsilon_{\parallel,m}}n(n+1) + \frac{1}{4}} - \frac{1}{2}.$$

It is seen that by changing the anisotropy, we can control the index of the Bessel function for the electric modes.

To model the graded-index profile, we have discretized the metamaterial shell to 40 homogenous layers. By applying the boundary conditions at the interface between each layer, we can find the scattering coefficients of the particle with metamaterial shell

### 4.2 Scattering coefficients of electric and magnetic multipoles in the core and in meta-shell

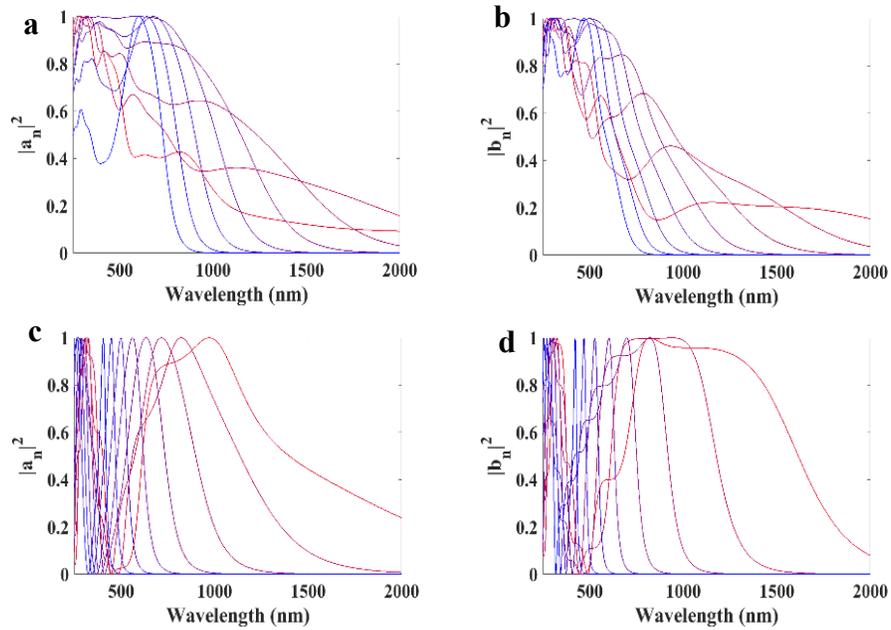

**Figure S9. Scattering coefficients of electric and magnetic multipoles in the core sphere and in the meta-shell.** Scattering coefficients of (**a**) electric mode and (**b**) magnetic mode in the meta-shell, (**c**) electric mode and (**d**) magnetic mode in the SiO2 core μ-sphere, d = 1 μm.

## 4.3 Mie resonance engineering with radial graded index and spherical anisotropy

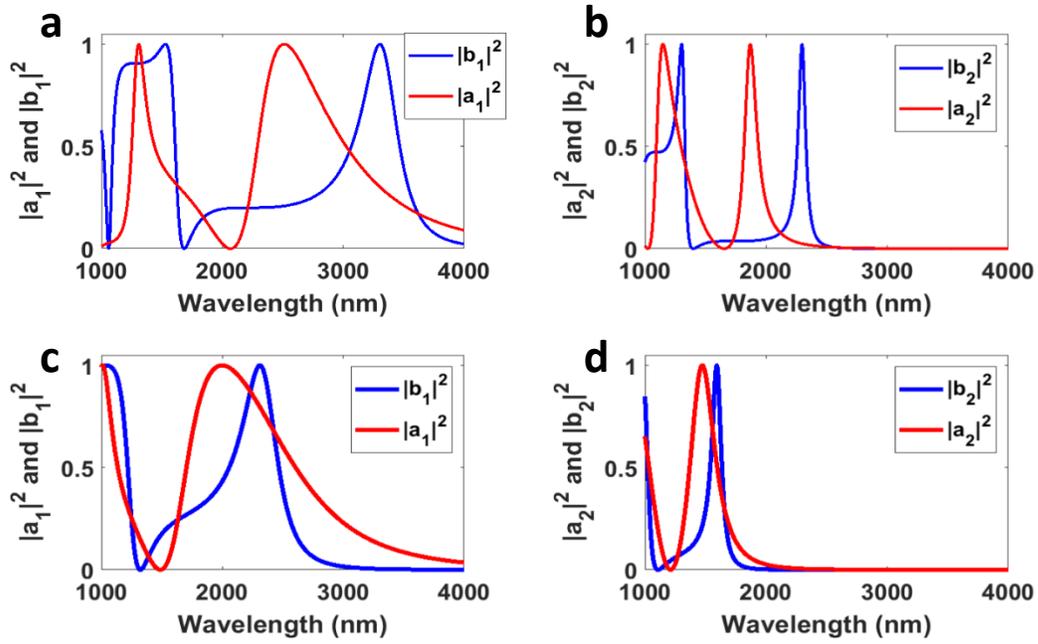

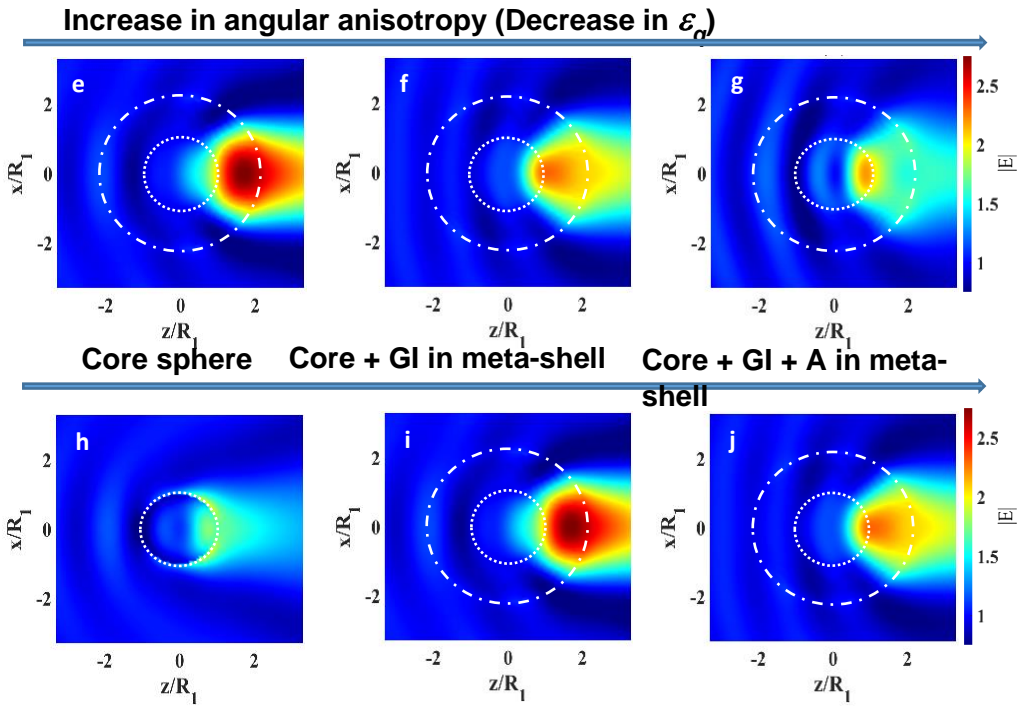

**Figure S10. Mie resonance engineering the meta-shell.** The diameter of the core sphere is $d_{core}$ = 1 mm, thickness of the meta-shell is $t_{shell}$ = 600 nm. The permittivity of the sphere is $e_{core}$ = 2.2; **a-d**, Increase in the radial graded index increases the spectral overlap between the electric and magnetic modes. Scattering coefficients of **a**) ED and MD, **b**) EQ and MQ for an isotropic sphere of $d$ = 1.4 μm, $\varepsilon$ = 10. Scattering coefficients of **c**) ED and MD, **d**) EQ and MQ for a core shell model ascribed with graded index profile in the shell. Here, core $d$ = 1 μm, $\varepsilon$ = 10 and shell $d$ = 0.4 μm and $\varepsilon$ radially decrease from 10 (core) to 1 (shell outer peripheries); **e-g**, increase in the anisotropy shifts the hotspot towards the core sphere interface, (**e**) $e_{q,\,shell}$ : 2.2, (**f**) $e_{q,\,shell}$ : 1.5, (**g**) $e_{q,\,shell}$ : 1; Here, the radial anisotropy is kept constant, $e_{r,\,shell}$ = 2.2. The input wavelength is $\lambda$ = 1550 nm (the input wavelength is $\lambda$ = 900 nm for identical plots in the main text, **Figure 3**); **h-j**, matching the analytical parameters to empirical measurements, photonic nanojet features are reproduced for (**h**) the core sphere and (**j**) the model MSP2.2 observed in the FDTD full wave simulation (**SI 3.2**); (**i**) While the peak $|E|$ is higher without the spherical anisotropy in the meta-shell, the region of peak $|E|$ shift towards, but exterior to, the core interface, at which the density of the $\chi^{(2)}$ nanostructures are at its highest.

**Comment:** This section shows further example of how increase in the radial gradient index helps with the spectral overlap of electric and magnetic modes. Compared to an isotropic sphere ($d$ = 1.4 μm, $\varepsilon$ = 10), there is significant increase in the spectral overlap between the electric and magnetic modes for the core-shell model ($d$ = 1 μm, $\varepsilon$ = 10) in which the shell is ascribed with graded index profile ($d$ = 0.4 μm, $\varepsilon$ = 10 -> 1). **Figure S10a** shows scattering coefficients of ED and MD and **S10b** shows scattering coefficients of EQ and MQ for the isotropic sphere. **Figure S10c** shows scattering coefficients of ED and MD and **S10d** shows scattering coefficients of EQ and MQ for core-shell model with graded index in the shell.

## 5. FDTD simulations – Nonlinear optics with MSP

### 5.1 Normalized conversion efficiency $\eta_{SHG}$ ($W^{-1}$) calculations

**Comment:** In running the full wave FDTD (Lumerical FDTD Solutions) calculations for second harmonic processes, the collection wavelengths $\lambda_{2\omega}$ is set to the second harmonic ranges and the simulation is carried out in a "nonorm" state. This removes numerical artifacts associated with the continuous wave approximation/normalization from the input pulse, $\lambda_\omega$, whose pulse length $\tau_\omega$ does not expand to the collection wavelength $\lambda_{2\omega}$. The output power, $\widetilde{P_{2\omega}}$, collected from the frequency domain field and power monitor is separately renormalized with the output pulse length $\tau_{2\omega}$, to obtain $P_{2\omega}$. The input power, $\widetilde{P_\omega}$, collected from the frequency domain monitor is renormalized with the pulse length $\tau_\omega$. This represents the peak power, $P_\omega$, delivered to the model MSP.

$$P_{2\omega} = \frac{\widetilde{P_{2\omega}}}{\tau_{2\omega}^2}$$

$$P_\omega = \frac{\widetilde{P_\omega}}{\tau_\omega^2}$$

The normalized SHG conversion efficiency, $\eta_{SHG}$, is obtained by dividing the output power by the input power squared.

$$\eta_{SHG} = \frac{P_{2\omega}}{P_\omega^2} = \frac{\widetilde{P_{2\omega}}}{\widetilde{P_\omega}^2} \times \frac{\tau_\omega^4}{\tau_{2\omega}^2}$$

## 5.2 Effect of core refractive index on the SHG conversion efficiency

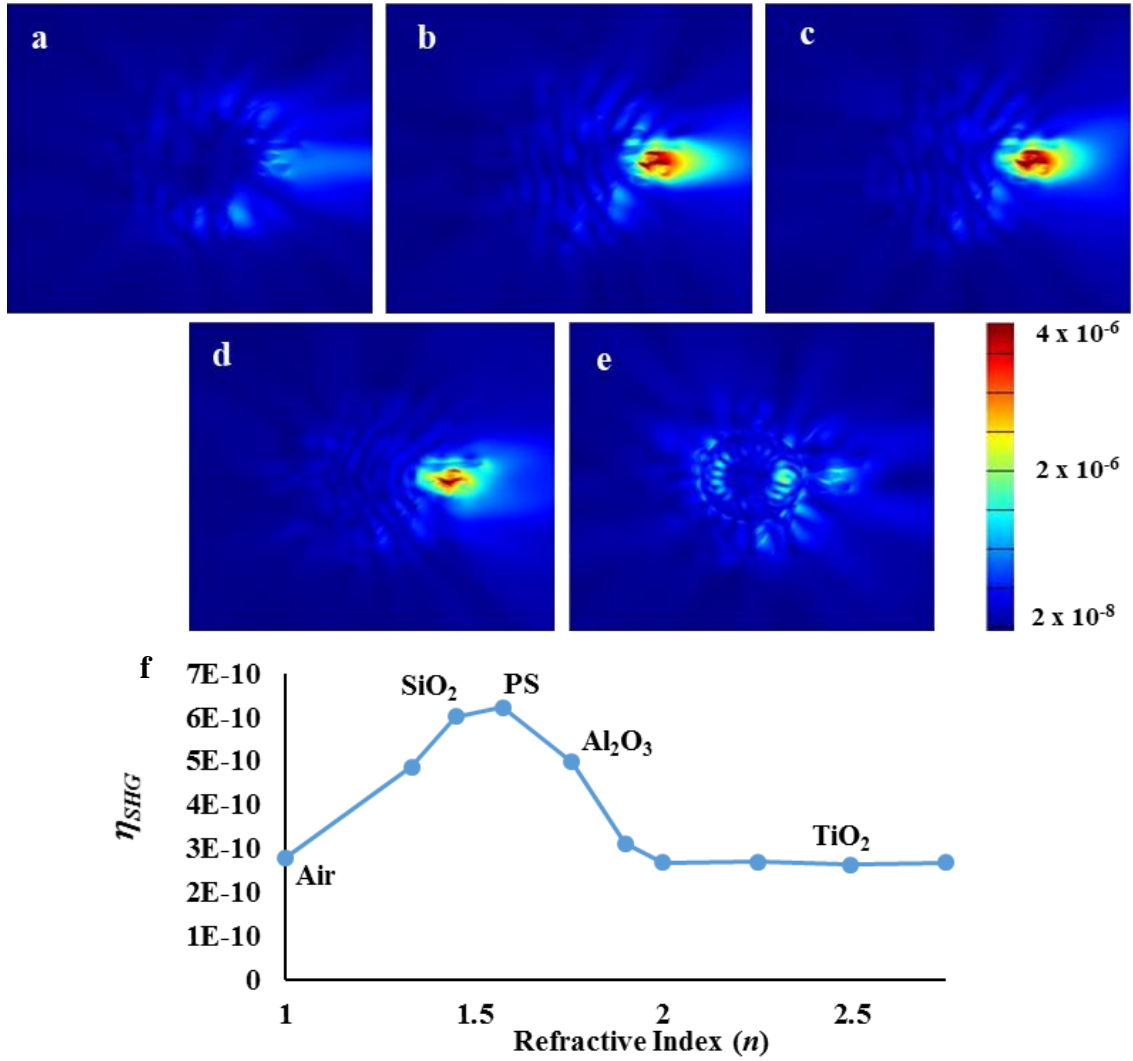

**Figure S11. Influence of core refractive index $n$ on the SHG conversion efficiency $\eta_{SHG}$ of the MSP.**
FDTD full wave simulation displaying the middle cross-sectional electric field profile, |E|, of the SHG generated by the MSP having (**a**) no core and (**b**) $SiO_2$ ($n$ = 1.47), (**c**) PS ($n$ = 1.57), (**d**) $Al2O_3$ ($n$ = 176), (**e**) $TiO_2$ ($n$ = 2.5) core sphere ($d$ = 1 μm). The thickness of the meta-shell is 600 nm. (**f**) the second harmonic conversion efficiency $\eta_{SHG}$ reaches the maximum with increase in the photonic nanojet hotspot |E| and when its spatial profile is fully exterior to the core sphere but closest to the core and meta-shell interface, at which the $\chi^{(2)}$ nanostructures are at its highest density.

**Comment 1:** The FDTD simulation of the normalized SHG conversion efficiency, $\eta_{SHG}$ ($W^{-1}$), shows that $\eta_{SHG}$ is highest for the MSP having a PS or a SiO$_2$ core. This is an expected outcome and can be predicted from the photonic nanojet features shown in **Section 3.3**. Here, the region of peak |E| in the nanojet hotspot is fully exterior to the core sphere, and at the same time, closest to the core interface at which the $\chi^{(2)}$ nanostructures are at its highest densities. With increasing $n$, the hotspots gradually internalize that leads to a decrease in the $\eta_{SHG}$. From $n > 2$, the photonic nanojet diminishes and hotspots are mostly confined within the core sphere and is accompanied by significant reduction in the $\eta_{SHG}$.

**Comment 2:** In calculating the conversion efficiency, we simplified the optical nonlinearities of ZnO NR to have isotropic susceptibility tensor having an effective $\chi^{(2)} = 15$ pm/V. The reference 13 reports a range of nonlinear coefficients from $d_{eff} = 2$ pm/V to 15 pm/V [13,14]. We have chosen a median value ($d_{eff} = 7.5$ pm/V) from the reported range to numerically approximate second harmonic conversion of the MSP.

# 6. Experiment – Confocal Microscopy

## 6.1 SHG by MSP observed with confocal microscopy

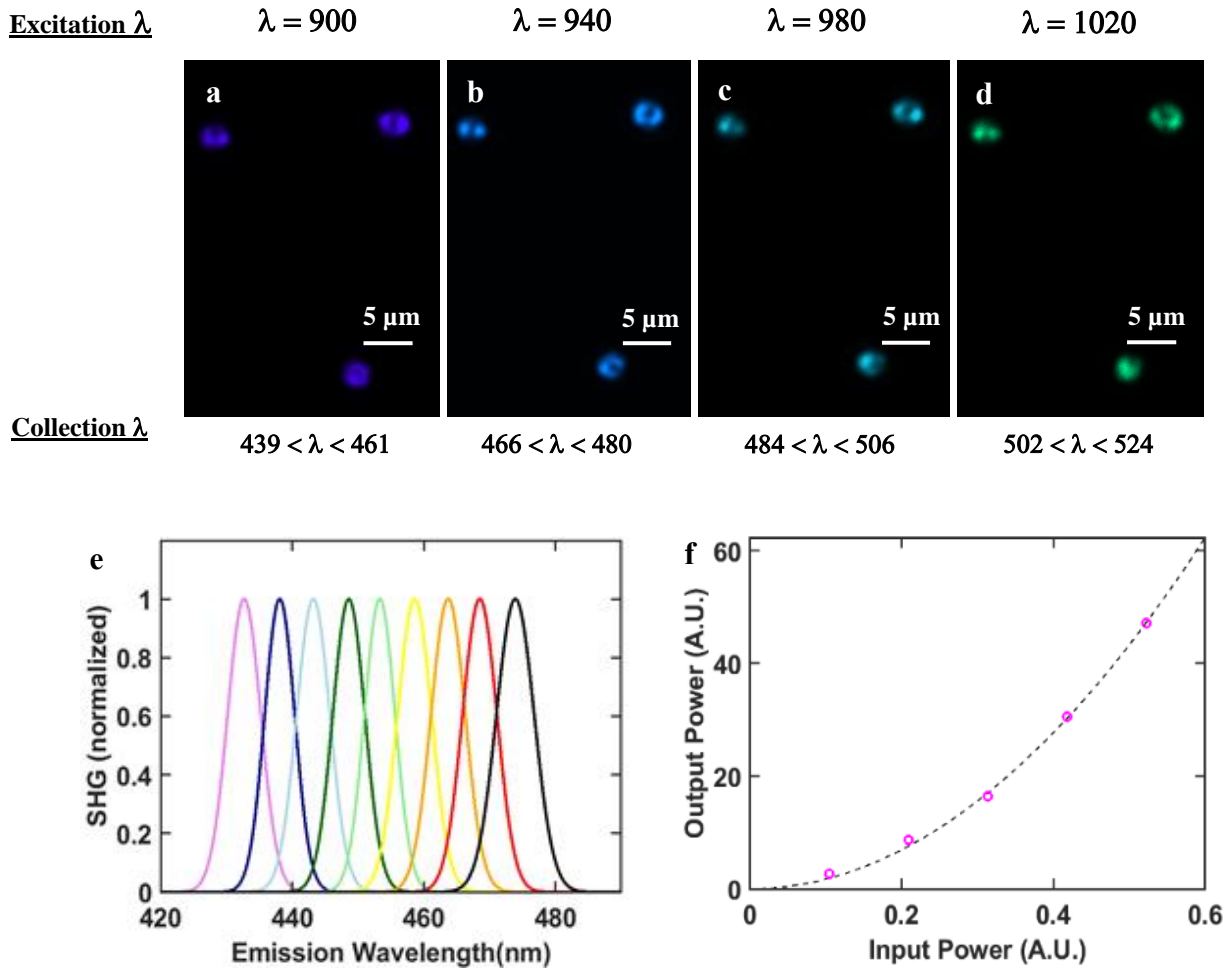

**Figure S12. Second harmonic generation of light by the MSP.** The MSP employed here are from **Figure 1d** in the main text, $d_{core}$ = 1 µm, $l$ = 600 nm, $w$ = 120 nm (designated MSP 2.2). **a-f**, experimental observation confirming SHG of light by the MSP 2.2; images observed with confocal microscopy when excited with 2-photon laser at input wavelengths (**a**) λ = 900 nm, (**b**) λ = 940 nm, (**c**) λ = 980 nm, (**d**) λ = 1020 nm and collected at detector ranges of 439 nm < λ < 461 nm, 466 nm < λ < 480 nm, 484 nm < λ < 506 nm, 502 nm < λ < 524 nm, respectively, (**e**) spectral collection from the MSP2.2 upon irradiation at 860 nm < λ < 960 nm at 10 nm increments, (**f**) the SHG output power (magenta, circle) demonstrates quadratic dependence to the input power (black, perforated, fit to y = $ax^2$, R-square: 0.9983);

**Comment:** Low numerical aperture objective (*NA* = 0.16) was utilized to ensure complete enclosure of MSP2.2 within the spot size (3.53 µm).

**6.2 SHG is not detectable from individual/small aggregate of ZnO NRs at identical settings.**

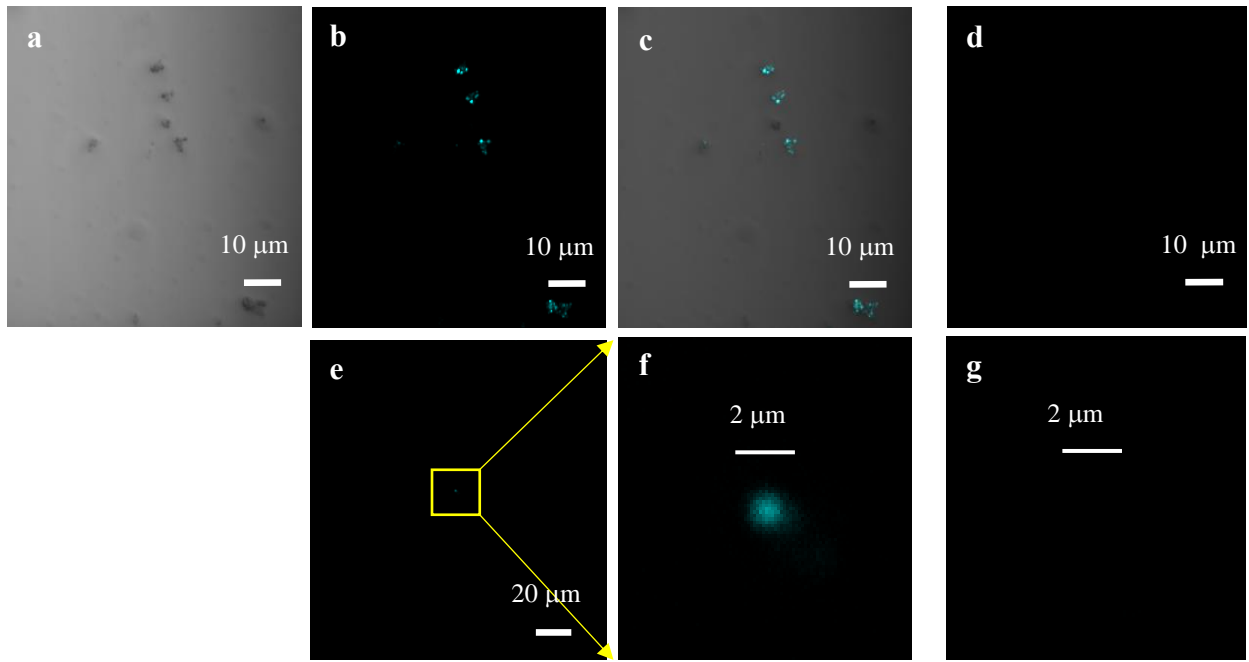

**Figure S13. SHG by small aggregates of ZnO NRs. a-d,** confocal microscopy images of small aggregates of ZnO NRs; (**a**) Transmission images of small aggregates of ZnO NRs, (**b**) The SHG images of small aggregates of ZnO NRs, irradiated at λ = 860 nm at maximum available power. The detectors are set to collect signals at 410 nm < $\lambda_{output}$ < 464 nm. (**c**) the overlay of the SHG signals on the transmission image, (**d**) at identical imaging conditions that was utilized to characterize the SHG of the MSP2.2 (irradiated at 5% of the maximum available power through 5X objective), no visible signals were detected from the ZnO NR aggregates. A 50X objective was utilized to irradiate, collect and image the ZnO NRs in (**a-d**); **e-g,** confocal microscopy images of the SHG by the ZnO NRs that may be either a very small aggregate or an individual NR. The difference can not be distinguished due to diffraction limitations, (**e**) The SHG images obtained when irradiated with 2-photon laser at input wavelengths λ = 860 nm at maximum available power through a 100X objective. The signals are collected at 410 nm < $\lambda_{output}$ < 464 nm, (**f**) 3X magnified image of (**e**), (**g**) At identical imaging conditions used for detecting and characterizing SHG from the MSP2.2, no visible signals were detected.

## 6.3 Expected enhancement in the conversion efficiency: MSP2.2 vs. ZnO NR

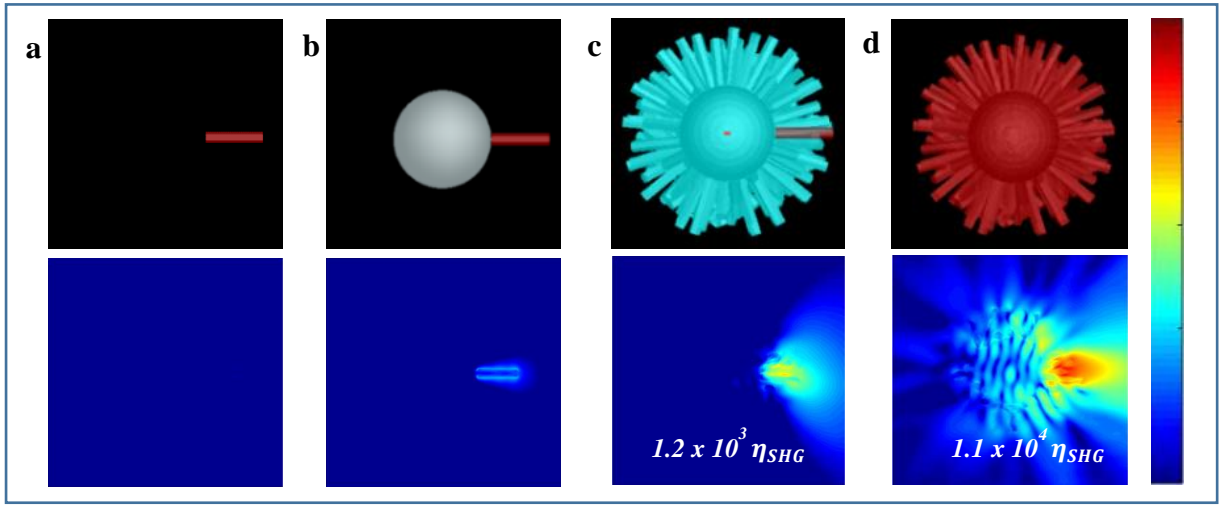

**Figure S14. a-d**, FDTD simulation demonstrating 4 orders of magnitude enhancement in the SHG conversion efficiency $\eta_{SHG}$ between an MSP2.2 and a ZnO NR with pulsed input centered at λ = 900 nm. The heat map depicts |E| of the SHG generated by the MSP. **e-f**, numerical artifact exist when carrying out |E| profile calculation in the Lumerical FDTD. This is more apparent with pulsed input centered at λ = 900 nm. The numerical artifact can be reduced by varying the mesh sizes, (**e**) 15 nm, (**f**) 20 nm. On the other hand, the magnitude of the numerical artifact oscillates with mesh sizes. On the other hand, there is only minor deviation in the calculation results.

**Comment 1**: We also estimated the expected enhancement in the normalized conversion efficiency, $\eta_{SHG}$, from a ZnO NR to an MSP2.2 for a pulsed input centered at λ = 900 nm (τ = 100 fs), which depicts the confocal microscopy configuration used for initial characterization of the SHG by the MSP. When placing a single $\chi^{(2)}$ ZnO nanorod at the shadow side of a silica core, the photonic jet enhances its second harmonic $\eta_{SHG}$ by an approximate 34 fold. In the presence of a meta-shell that features the effective optical responses, but without the $\chi^{(2)}$, there is additional 36-fold increase in the $\eta_{SHG}$. Finally, when 2$^{nd}$ order nonlinearity is assigned to the meta-shell in its entirety, as is the case with the MSP, there is an additional 9-fold increase in the $\eta_{SHG}$. Hence, there is a 4 order of magnitude enhancement in the $\eta_{SHG}$ between a ZnO NR and their assemblies into an MSP.

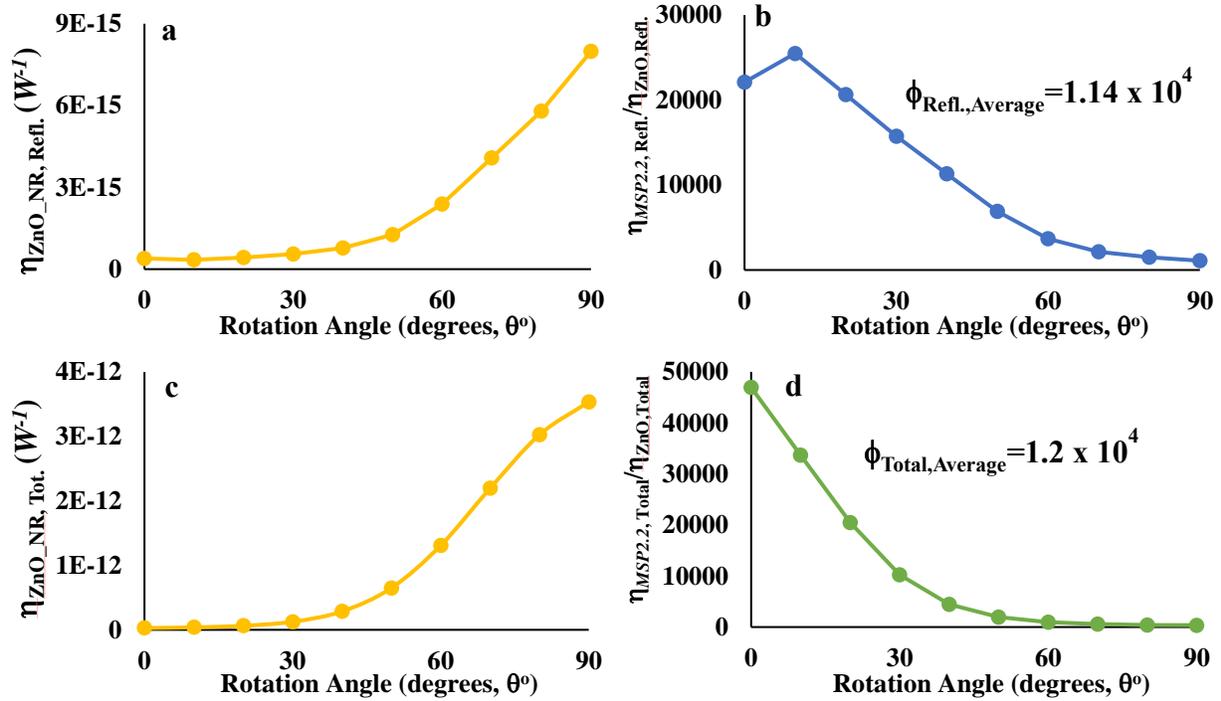

**Figure S15. Full wave FDTD calculations of the SHG $\eta_{SHG}$ of ZnO NRs and the model MSP2.2.** The normalized conversion efficiency of ZnO NR ($l$ = 600 nm, $w$ = 120 nm) are calculated at $10^o$ increments in their orientations. The orientation of the ZnO NR is at $0^o$ when the polarization of the incident light is perpendicular to the ZnO NR $c$-axis, at $90^o$ when the polarization of the incident light is aligned with the ZnO NR c-axis. **a-b**, orientation dependent normalized conversion efficiency of ZnO calculated from the output collected in the *reflection monitor* in the simulation. The FDTD simulation emulates the confocal microscopy settings (NA = 0.16). (**a**) angle dependent normalized conversion efficiency for the ZnO NR, $\eta_{ZnO-NR, Refl.}$; (**b**) angle dependent enhancement in the normalized conversion efficiency by the model MSP2.2 from a single ZnO NR, $\phi_{Refl.} = \eta_{MSP2.2, Refl.}/\eta_{ZnO-NR,Refl.}$. The normalized conversion efficiency of the model MSP2.2 in the reflection monitor is $\eta_{MSP2.2, Refl.} = 8.72 \times 10^{-12}$ W$^{-1}$, and the conversion enhancement averaged over all angles is $\phi_{Refl.,Average} = \eta_{MSP2.2, Refl.,Average}/\eta_{ZnO-NR,Refl.,Average} = 1.11 \times 10^4$. **c-d**, FDTD simulation of orientation dependent normalized conversion efficiency calculated from the *total output* of the SHG; (**c**) angle dependent normalized conversion efficiency for the ZnO NR, $\eta_{ZnO-NR,Total}$; (**d**) angle dependent enhancement in the normalized conversion efficiency by the model MSP2.2 from a ZnO NR, $\phi_{Total.} = \eta_{MSP2.2, Total.}/\eta_{ZnO-NR,Total}$. The normalized conversion efficiency of the model MSP2.2 calculated from the total SHG output is $\eta_{MSP2.2, Total} = 1.27 \times 10^{-9}$ W$^{-1}$, and the conversion enhancement averaged over all angles is $\phi_{Total,Average} = \eta_{MSP2.2, Total,Average}/\eta_{ZnO-NR,Total,Average} = 1.20 \times 10^4$. A TFSF pulse of $\tau$ = 100 fs centered at an arbitrary wavelength of $\lambda$ = 896 nm was used as an input pump in the FDTD simulation that was applied to both the ZnO NRs and the MSP2.2.

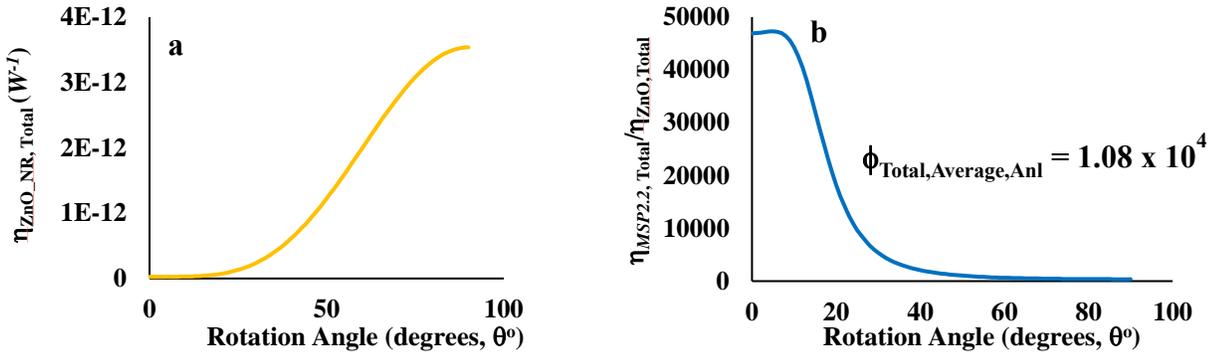

**Figure S16. Analytical calculations of the SHG conversion efficiency for ZnO NRs and the model MSP2.2.** Estimated enhancement in the normalized conversion efficiency by the MSP2.2 from a ZnO NR via analytical approach. (**a**) angle dependent normalized conversion efficiency for the ZnO NR, $\eta_{ZnO\_NR}$; (**b**) angle dependent enhancement in the normalized conversion efficiency by the model MSP2.2 from a single ZnO NR, $\phi_{Total.} = \eta_{MSP2.2, Total.}/\eta_{ZnO,Total}$. The conversion enhancement averaged over all angles is $\phi_{Total,Average,Anl} = \eta_{MSP2.2, Total,Average}/\eta_{ZnO,Total,Average,Anl} = 1.08 \times 10^4$.

**Comment:** In this section, FDTD simulation is employed to estimate the enhancement in the normalized SHG conversion efficiency by the MSP2.2 ($\eta_{MSP2.2}$) from a single ZnO NR ($\eta_{ZnO-NR}$), $\phi = \eta_{MSP2.2}/\eta_{ZnO-NR}$. The $\eta_{ZnO-NR}$ depends on its orientation relative to the polarization of the input pulse. The polarization of the input pulse is perpendicular to the ZnO NR *c*-axis at $0^o$ orientation and aligned at $90^o$ orientation. In the FDTD simulation that compares the conversion enhancement between the model MSP2.2 and a ZnO NR in the main text (**Figure 4c-f**) and **Figure S14,** only one orientation is considered. For a more comprehensive evaluation, the conversion enhancement ($\phi$) is evaluated over $0°$ to $90°$ in the ZnO NR orientation, at $10°$ increments, and then averaged, $\phi_{Average} = \eta_{MSP2.2}/\eta_{ZnO-NR}$. The $\eta_{ZnO-NR}$ and $\phi_{Average}$ are computed via three approaches: 1) FDTD calculation of the enhancement in the normalized conversion efficiency taking into account only the reflected SHG output ($\phi_{Refl.,Average}$), **Figure S15a, b**. This FDTD setup emulates the confocal microscopy setting (NA = 0.16) employed in section **S6.1** and

**S6.2, Figure S18**; 2) FDTD simulation of the enhancement in the normalized conversion efficiency taking into account the total SHG output ($\phi_{Total,Average}$), **Figure S15c, d**; 3) analytical approach to estimate the enhancement in the conversion efficiency ($\phi_{Total,Average,Anl}$) based on the following expression, **Figure S16**:

$$P_{2\omega} = \eta(\theta) \times P_\omega^2$$

$$\eta(\theta) = \eta_x P_\omega^2 \cos^4\theta + \eta_y P_\omega^2 \sin^4\theta$$

, where $\eta_x$ and $\eta_y$ are the normalized conversion efficiency of ZnO NR at 0° and 90° orientation respectively and are obtained from the FDTD simulations.

Due to spherical configuration, the normalized SHG conversion efficiency of the model MSP2.2, $\eta_{MSP2.2}$, are assumed orientation independent and are therefore evaluated at only one incident angle. The enhancement in the conversion efficiency by the MSP2.2 from a single ZnO NR averaged over a quadrant of angles are equivalent in all the three approaches considered; $\phi_{Refl.,Average} = 1.14 \times 10^4$, $\phi_{Total,Average} = 1.2 \times 10^4$, $\phi_{Total,Average,Anl} = 1.08 \times 10^4$.

## 6.4 Comparison of SHG conversion efficiency between MSP of different meta-shell thickness

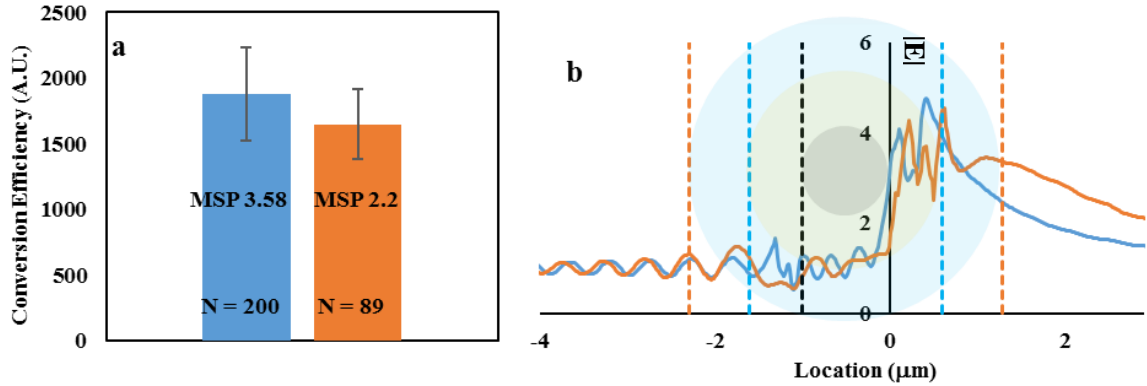

**Figure S17. Normalized conversion efficiency comparison between MSP2.2 and MSP3.58.** The MSP3.58 consists of $SiO_2$ core sphere ($d = 1$ μm) with meta-shell of thickness 1.29 μm (ZnO NR $l = 1290$ nm, $w = 120$). (**a**) SHG conversion efficiency, in arbitration units, obtained for the MSP3.58 ($N = 89$) and the MSP2.2 ($N = 200$). (**b**) FDTD simulation shows that the peak |E| at the photonic nanojet hotspot in the MSP3.58 and in the MSP2.2 are comparable. The blue and the orange line represents the |E| for MSP3.58 and MSP2.2, respectively.

**Comment 1:** The MSP3.58 consists of $SiO_2$ core sphere ($d = 1$ μm) with a meta-shell thickness of 1.29 μm (ZnO NR $l = 1290$ nm, $w = 120$). Interestingly, the normalized SHG conversion efficiency for the MSP3.58 ($\eta_{MSP3.58}$) and that of MSP2.2 ($\eta_{MSP2.2}$) are comparable ($\eta_{MSP3.58}/\eta_{MSP2.2} = 1.38$). This is due to comparable peak electric field |E| in the nanojet hotspot for the both types. The measurement was carried out with confocal microscopy in which the MSP are irradiated with femtosecond pulses (140fs, 80MHz) centered at $\lambda = 900$ nm and at 5% of its maximum available power. The images are collected at 439 nm $< \lambda_{output} <$ 461 nm. In order to fully immerse the MSP within the spotsize, a low *NA* (0.16) 5X objective was use. The SHG intensities obtained from the images are subtracted by the background noise, normalized by the gain settings, followed by normalization with the square of the input power reading.

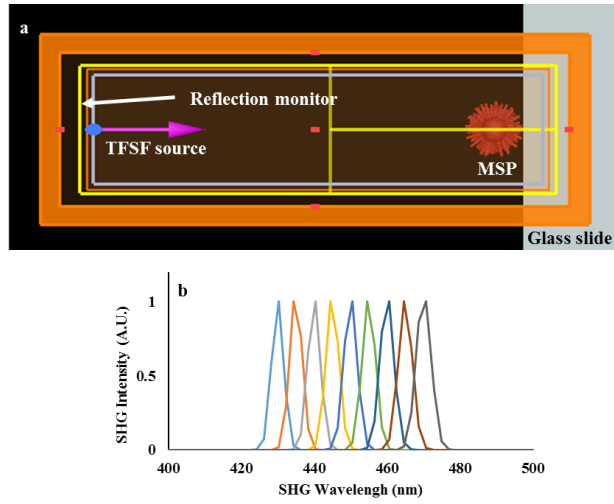

**Figure S18.** (**a**) FDTD simulation setup, *NA* = 0.16; (**b**) FDTD simulations of the SHG output from the MSP2.2 irradiated at different wavelengths ranging from 860 nm to 940 nm.

**Comment 2:** The geometry of the reflection monitor and its distance from the particle, in the FDTD simulation setup, is arranged to emulate the numerical aperture *NA* (0.16) of the objective employed in the confocal microscopy measurement, **Figure S18**. TFSF sources are employed as the input pulse centered at $\lambda_{input}$ = 900 nm with pulse length $\tau$ = 100 fs. A glass slide is placed behind the model MSP2.2 to recapture the experimental setting. The ZnO NRs are imparted with 2$^{nd}$ order nonlinear susceptibility $\chi^{(2)}_{eff}$ = 15 pm/V.

**Comment 3**: The FDTD simulation is carried out to estimate and compare the normalized conversion efficiency of the MSP.358 and MSP2.2. In accordance with the experimental setting, only the SHG output collected in the reflection monitor (NA = 0.16) is taken into account in the calculation, as in **comment 1**. The FDTD simulation also results in a similar normalized conversion efficiency between the MSP3.58 and the MSP2.2. While the total conversion efficiency for the MSP3.58 is $\eta_{\text{MSP3.58\_FDTD\_tot}} = 1.87 * 10^{-9} \text{ W}^{-1}$, the conversion efficiency evaluated with the reflected SHG is $\eta_{MSP3.58\_FDTD\_refl} = 8.81 * 10^{-12}$ W$^{-1}$. In the case of MSP2.2, while the total conversion efficiency is $\eta_{MSP3.58\_FDTD\_tot} = 1.23 * 10^{-9}$ W$^{-1}$, the conversion efficiency evaluated with the reflected SHG is $\eta_{MSP3.58\_FDTD\_refl} = 7.15 * 10^{-12}$ W$^{-1}$. Hence, the ratio of the conversion efficiency taking into account only the reflected SHG, as in the experimental setting, $\eta_{MSP3.58\_FDTD\_refl} / \eta_{MSP2.2\_FDTD\_refl} = 1.23$, which closely approximates the experimental findings.

## 6.5 Comparison of SHG conversion efficiency of MSP2.2 having SiO$_2$ core and TiO2 core.

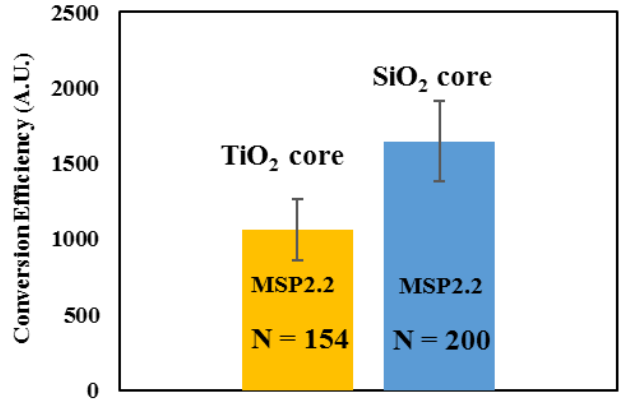

**Figure S19. Normalized conversion efficiency comparison between MSP2.2 having a SiO$_2$ core and a TiO$_2$ core.** Confocal microscopy is used to evaluate and compare the conversion efficiency between the two MSP2.2 types. See **Section 6.4**, **Comment 1** for the experimental measurement setting.

**Comment:** The SHG conversion efficiency by two MSP2.2 types, one having a SiO$_2$ core and one having a TiO$_2$ core are compared.

Due to the high refractive index of $TiO_2$ ($n = 2.5$ at $\lambda = 900$ nm) the photonic nanojet at the shadow side disappears and the hotspots are dispersed within and along the interface of the core sphere, as seen in the FDTD simulations in section **3.3**, **Figure S8**. This results in a lack of spatial overlap between the |E| hotspot and the $\chi^{(2)}$ nanostructures. Hence, reduction in the SHG conversion efficiency for the MSP2.2 with the $TiO_2$ core is expected. Such is observed through experimental measurement via confocal microscopy. The conversion efficiency ratio between MSP2.2 of $TiO_2$ and of $SiO_2$ core from the measurement is $\eta_{MSP2.2\_TiO2\_exp} / \eta_{MSP2.2\_SiO2\_exp} = 0.64$. The experimental setup is identical to that employed in section **6.1 – 6.4**. The FDTD simulation also results in a similar observation. While the total conversion efficiency for the MSP2.2 with TiO2 core is $\eta_{MSP2.2\_TiO2\_FDTD\_tot} = 4.7 * 10^{-10}$ $W^{-1}$, the conversion efficiency evaluated from the reflected SHG is $\eta_{MSP2.2\_TiO2\_FDTD\_refl} = 4.1 * 10^{-12}$ $W^{-1}$. The ratio of the conversion efficiency taking only the reflected SHG into consideration is $\eta_{MSP2.2\_TiO2\_FDTD\_refl} / \eta_{MSP2.2\_SiO2\_FDTD\_refl} = 0.57$, which closely approximates the experimental findings.

# 7 Experiment – Nonlinear optical microscopy

## 7.1 Schematics – NLO microscopy

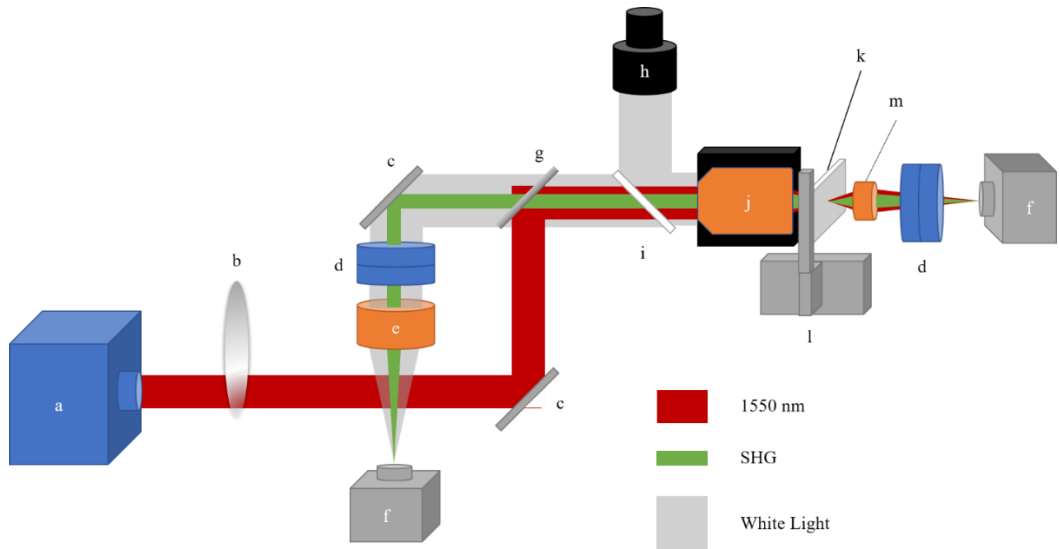

**Figure S20.** Schematic of the NLO microscopy setup used to measure the SHG generated by the MSP.

**Comment:** The schematic in **Figure S20** depicts the nonlinear optical (NLO) microscopy setup constructed to detect the SHG generated by the MSP. **Table S1** shows the optical components employed in the setup. Ultra-short pulse frequency comb (Menlo Systems, $\tau = 80$ fs , $f_{rep} = 250$ MHz) centered at $\lambda_{in} = 1550$ nm is used as the input source. The input pulse is subject to a continuously variable neutral density filter wheel to obtain variable input average power. The input beam is guided into a 50x objective (Mitutoyo Plan APO NIR, $NA = 0.42$) and focused onto a single particle on the sample slide with a spot size $d_f \approx 2.6$ µm. A collimated LED white light source was incorporated to locate and focus the input pulse into individual MSP. The backscattered SHG is subject to a pair of filters to remove the input light and the third harmonic generation. A 1" VIS lens was used to focus the SHG onto a photodiode power sensor. The total average power of the backscattered SHG is then determined by accounting for the power loss through each optical component. The transmission values of each optical component that the SHG passes through are shown in **Table S2**.

The forward scattered SHG was collected via aspheric condenser lens *(NA .79, F = 16 mm)* and subject to a pair of filters to remove the input pump and the higher order harmonic generations identical to the reflection mode setup. The total average power of the forward scattered SHG is then determined by accounting for the power loss through each optical component, listed in **Table S2**. The aspheric condenser lens was replaced with a 60X objective (Nikon, M Plan 60, 630770) when taking the forward scattered images to obtain the farfield pattern.

| a | Frequency comb centered at 1550 nm, 80fs, 250 MHz |
|---|---|
| b | Round continuously variable ND filter wheel |
| c | Silver mirrors |
| d | Filters to remove pump and higher order harmonic wavelengths |
| e | VIS lens |
| f | Camera/780 nm/ detector |
| g | Shortpass dichroic mirror |
| h | Collimated LED white light source |
| i | VIS beam splitter |
| j | 50x Objective |
| k | Sample slide coated with meta-shell supraparticles |
| l | x, y-stage |
| m | Aspheric condenser lens / 60 X objective |

**Table S1**: List of labeled components in **Figure S20**

| Optical Component | Percent Transmission (%) | |
|---|---|---|
| 50x Mitutoyo Plan APO NIR, NA = 0.42 | 67 (**Supplementary Information section 7.2)** | Emperical |
| Thorlabs DMSP950 | 97.6979 | Thorlabs Raw Data |
| Thorlabs FESH0800 | 98.201267 | Thorlabs Raw Data |
| Thorlabs FELH0550 | 94.94641 | Thorlabs Raw Data |
| Thorlabs VIS (A-coating) lenses | 96.56032 | Thorlabs Raw Data |

**Table S2**: Percent transmission values for various optical components in the experimental setup that were used to determine the total average SHG power generated. The values are for light of wavelength $\lambda = 780$ nm.

## 7.2 Schematics – Objective calibration

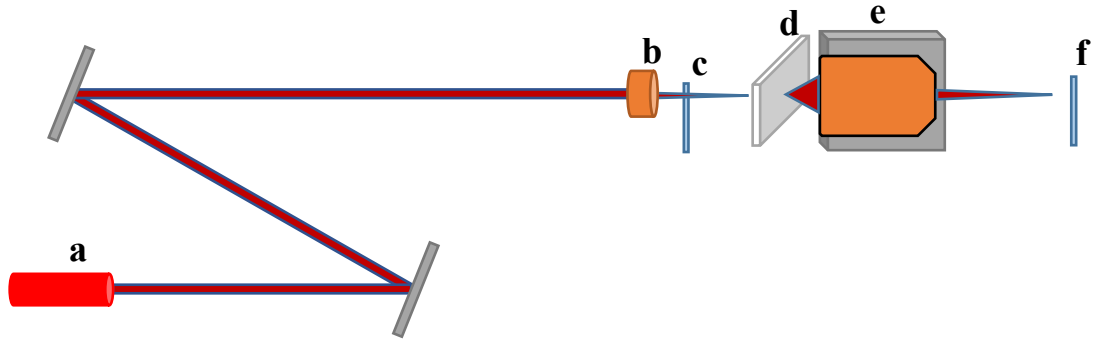

**Figure S21.** Schematic for the objective calibration setup

| a | Laser diode, $\lambda$ = 775 nm |
| b | Lens, $f$ = 15 mm |
| c | Power detector |
| d | Glass slide |
| e | Objective to be calibrated |
| f | Power detector |

**Table S3**: List of labeled components in **Figure S21**

## 7.3 Calibration - Average input power delivered to the sample

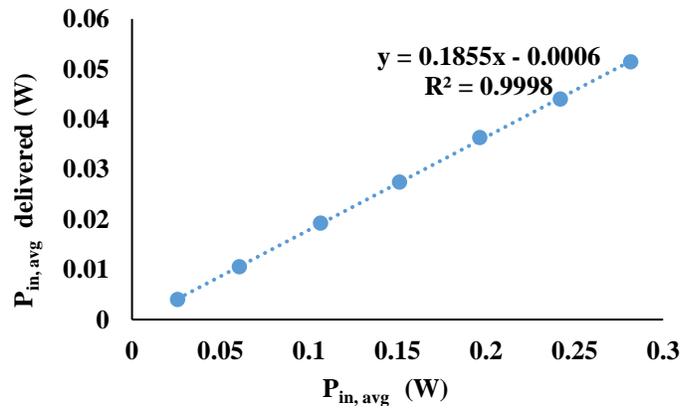

**Figure S22.** Calibration curve for the average input power, $P_{in,avg}$, delivered to the sample.

**Comment:** The calibration curve for the average input power delivered to the sample is obtained from the power reading of the input source recorded prior to the continuously variable neutral density filter wheel and that immediately after the objective. Thermal power sensor head (Thorlabs) is utilized for the measurement.

## 7.4 FDTD - Simulation setup to replicate NLO microscopy setting

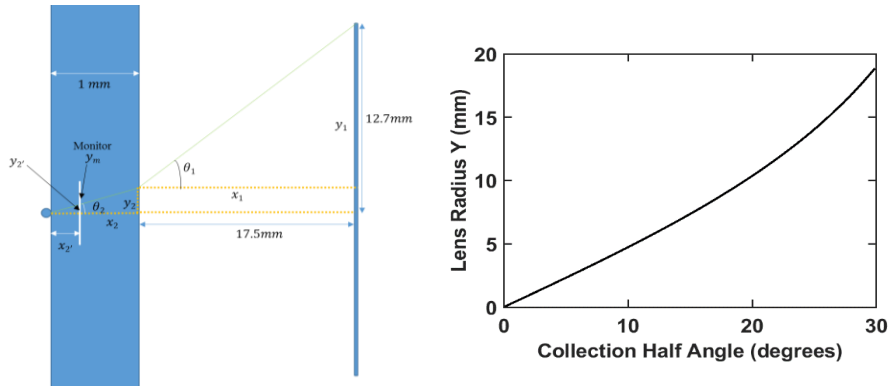

**Figure S23.** Schematic for the objective calibration setup

**Comment 1:** TFSF sources are employed as the input pulse centered at $\lambda_{input} = 1550$ nm with pulse length $\tau = 80$ fs. A glass slide is placed behind the model MSP2.2 to recapture the experimental setting. The ZnO NRs are imparted with 2$^{nd}$ order nonlinear susceptibility $\chi^{(2)}_{eff} = 15$ pm/V.

**Comment2:** The power monitor in the FDTD simulation is set up to mirror the half collection angle devised in the experimental setup. Taking $n = 1.5$ for the glass slide, the following relationship is satisfied

$$Y = y1 + y2 = \tan(\theta2) + 17.5 * \tan(\theta1)$$

$$= \tan(\theta2) + 17.5 * \tan(Sin^{-1}(1.5 * sin(\theta2))) < 12.7 \text{ mm}$$

, where Y is radius of the lens. From the relationship, the maximum half angle of collection for the transmission measurement of the SHG signal, using an aspherical condenser lens ($d = 25.4$ mm, $NA = 0.8$), is approximately 23º. The half collection angle for the reflection measurement is approximately 25º.

## 7.5 FDTD - Normalized Conversion Efficiency

FDTD simulation is set up to emulate the experimental settings in the NLO microscopy (**Section 7.4**). From the FDTD calculations, the normalized SHG conversion efficiency by the MSP2.2, where the simulation takes into account only the reflected SHG captured by the objective used in the experiment, is $\eta_{MSP2.2, Refl.} = 8.53 \times 10^{-12}$ $W^{-1}$. Taking into account only the forward scattered SHG captured by the aspheric condenser lens, the normalized SHG conversion efficiency by the MSP2.2 is $\eta_{MSP2.2, forward} = 6.8 \times 10^{-11}$ $W^{-1}$. These values are close to the experimental results observed from the custom built NLO microscopy. The normalized conversion efficiency for the total SHG generated by the MSP2.2 is $\eta_{MSP2.2, Total} = 1.48 \times 10^{-10}$ $W^{-1}$.

## 7.6 Farfield Radiation Pattern (Reflection) and quadratic dependence to input power

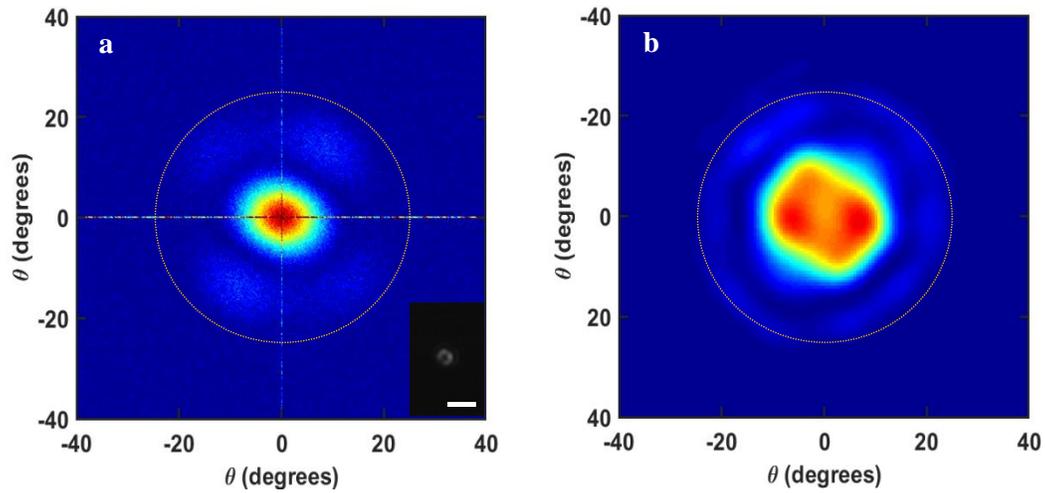

**Figure S24. Farfield radiation pattern of the reflected SHG.** (**a**) Far field radiation pattern of the reflected SHG (inset: image of the reflected SHG), (**b**) FDTD simulation of the far field radiation pattern of the reflected SHG. Dotted circle represents the angle of collection for the objective (*NA* 0.42) utilized in the imaging setup. scale bar: 5 μm

**Comment 1**: For imaging reflected SHG, the half angle of collection for the objective used in the setup is $\theta = 24.83^o \approx 25^o$. For imaging the forward scattered SHG, we replaced the aspheric condenser lens with the 60X objective (**Section 7.1**), increasing the half angle of collection to approximately $\theta = 27.81^o \approx 28^o$.

**Comment 2**: The far field radiation pattern is obtained by taking FFT of the SHG image in which the background noise is subtracted.

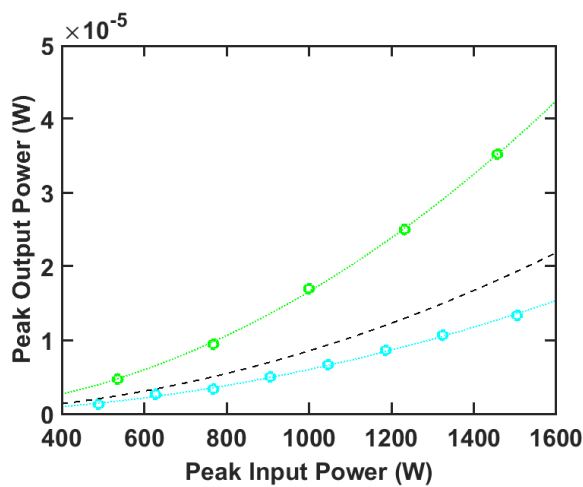

**Figure S25. SHG output power display quadratic dependence to the input power**. The samples displaying an upper limit (green circle) and a lower limit (cyan circle) in the conversion efficiency values are plotted. The dotted green line (R = 0.996, $\eta = 1.657 \times 10^{-11}$) and the dotted cyan line (R = 0.9997, $\eta = 5.995 \times 10^{-11}$) are the respective curve fit ($y = \eta x^2$) to the quadratic relationship. The quadratic relationship calculated from the FDTD simulation (red perforated line, $\eta = 8.53 \times 10^{-12}$) fall within the experimental range.

## 8. Reference (Supplementary Information)

**References (Main Text)**